\documentclass[a4paper,cite]{article}
\usepackage{latexsym}

\newcommand{\lbl}[1]{\label{eq:#1}}
\newcommand{ \rf}[1]{(\ref{eq:#1})}

\newcommand{\be}{\begin{equation}}
\newcommand{\ee}{\end{equation}}
\newcommand{\bea}{\begin{eqnarray}}
\newcommand{\eea}{\end{eqnarray}}
\newcommand{\setl}{\setlength\arraycolsep{2pt}}

\newcommand{\noi}{\noindent}
\newcommand{\nn}{\nonumber}
\newcommand{\ra}{\rightarrow}

\newcommand{\lesssim}{ {\
\lower-1.2pt\vbox{\hbox{\rlap{$<$}\lower5pt\vbox{\hbox{$\sim$}}}}\ }
}
\newcommand{\gtrsim}{ {\
\lower-1.2pt\vbox{\hbox{\rlap{$>$}\lower5pt\vbox{\hbox{$\sim$}}}}\ }
}

\newcommand{\cA}{{\cal A}}

\newcommand{\cC}{{\cal C}}

\newcommand{\cH}{{\cal H}}

\newcommand{\cL}{{\cal L}}
\newcommand{\cM}{{\cal M}}

\newcommand{\cO}{{\cal O}}
\newcommand{\cP}{{\cal P}}
\newcommand{\cR}{{\cal R}}

\newcommand{\cU}{{\cal U}}

\newcommand{\tr}{\mbox{\rm tr}}

\newcommand{\MeV}{\mbox{\rm MeV}}
\newcommand{\GeV}{\mbox{\rm GeV}}

\newcommand{\with}{\mbox{\rm with}}

\newcommand{\annd}{\mbox{\rm and}}
\newcommand{\foor}{\mbox{\rm for}}

\newcommand{\als}{\alpha_{\mbox{\rm {\scriptsize s}}}}

\newcommand{\GF}{G_{\mbox{\rm {\tiny F}}}}

\newcommand{\qs}{\not \! q}

\newcommand{\gL}{\frac{1-\gamma_{5}}{2}}
\newcommand{\gR}{\frac{1+\gamma_{5}}{2}}

\newcommand{\elm}{\mbox{\rm {\tiny em}}}
\newcommand{\nc}{\mbox{\rm {\tiny nc}}}

\newcommand{\ksls}{\not \! k}

\newcommand{\qsls}{\not \! q}
\newcommand{\pslsh}{\not \! p}

\newcommand{\stern}{\langle\bar{\psi}\psi\rangle}

\newcommand{\dd}{\!\cdot\!}

\input epsf

\def\theequation{\arabic{section}.\arabic{equation}}

\begin{document}

\begin{titlepage}

\begin{flushright} CPT-2001/P.4133
\end{flushright}
\vspace*{1.5cm}
\begin{center}
{\Large \bf Electroweak Hadronic Contributions to $g_{\mu}-2$}\\[3.0cm]

{\bf Marc~Knecht}$^a$, {\bf Santiago~Peris}$^b$, {\bf
Michel~Perrottet}$^a$  and {\bf Eduardo~de Rafael}$^a$\\[1cm]

$^a$  Centre  de Physique Th{\'e}orique\\
       CNRS-Luminy, Case 907\\
    F-13288 Marseille Cedex 9, France\\[0.5cm]
$^b$ Grup de F{\'\i}sica Te{\`o}rica and IFAE\\ Universitat
Aut{\`o}noma
de Barcelona, 08193 Barcelona, Spain.\\

\end{center}

\vspace*{1.0cm}

\begin{abstract}

We reanalyze the two-loop electroweak hadronic contributions to
the muon $g-2$ that may be enhanced by large logarithms. The present
evaluation is improved over those already existing in the literature by
the implementation of the current algebra Ward identities and  the
inclusion of the correct short--distance QCD behaviour of the relevant
hadronic Green's function.

\end{abstract}

\end{titlepage}

\section{\normalsize Introduction}
\lbl{int}

\noi
The latest result from the $g_{\mu}-2$ experiment at
BNL~\cite{BNL} reported by the E821 collaboration has triggered a renewal
of interest on the theoretical prediction of the anomalous
magnetic moment of the muon
$a_{\mu}\equiv
\frac{1}{2}(g_{\mu}-2)$ in the Standard Theory~\footnote{For a recent
review see e.g. ref.~\cite{CM01} where some of the earlier references
can
also be found.}. The attention
has been focused mostly on the hadronic vacuum polarization
contribution, particularly on the accuracy of its
determination~\cite{DH98,J00,N01,DTY01,CLS01}. The major theoretical
change, however, comes from a new evaluation of the dominant
pion--pole contribution to the hadronic light--by--light
scattering~\cite{KNb01}, which has unravelled a sign mistake in
earlier calculations~\cite{KNO85,HKS9596,BPP,BP01,bartos01}. The
correctness of the result in ref.~\cite{KNb01} was corroborated by a
renormalization group argument within the low--energy effective theory of
the Standard Theory in ref.~\cite{KNPdeR01}, with subsequent
developments in refs.~\cite{Ba01,BCM01} and \cite{RW02}. The previous
calculations have now been amended
correspondingly~\cite{HK01,BPPe01,bartosR01}.

Here we wish to report on a new calculation concerning an interesting
class of hadronic contributions which appear at the two--loop level in
the electroweak sector. They are the ones generated by the hadronic
$\gamma-\gamma-Z$ vertex, with one $\gamma$ and the
$Z$--boson attached to a muon line, as illustrated by the Feynman
diagrams in Fig.~1.

As first noticed in~\cite{KKSS92} these contributions are
particularly interesting since, in principle, they can be
enhanced by a large $\log(M_Z^2/m^2_{\mathrm{loop}})$ factor, where
$m_{\mathrm{loop}}$ is the relevant scale in the shaded loop in
Fig.~1. However, in the absence of the strong interactions, there is
an important cancellation between leptons and quarks within a
given family, as a consequence of the anomaly--free charge
assignment in the Standard Theory~\cite{PPdeR95,CKM95}. The
purpose of the present work is to analyze what is left out of this
cancellation in the sector of the $u,d$ and $s$ quarks, where
the strong interactions turn out to play a subtle role.

When dealing with the strong interaction effects, it has become a common
{\it simplifying practice} among some theorists, to assume that
the main effect of the strong
interactions is {\it dual} to a modification of the $u,d,s$ quark masses
in the QCD Lagrangian, to ``constituent--like'' quark masses
of the order of
$300\,\MeV$ for the
$u,d$ quarks and
$500\,\MeV$ for the $s$ quark. In this way the contributions
from Fig.~1 were found to be rather
small~\cite{PPdeR95,CKM95}~\footnote{Notice that
ref.~\cite{PPdeR95} uses a more sophisticated
version of this approach where at least chiral symmetry breaking
is correctly implemented~\cite{BBdeR93}.}. However,
in view of the expected accuracy which the new BNL experiment will
eventually reach,  it becomes necessary  to have a more reliable
determination of the size of these contributions. This has now become
possible within the framework of the $1/N_c$ expansion in QCD. There are
indeed recent theoretical developments of this non--perturbative analytic
approach, which have been applied mostly to the calculation of
hadronic weak processes~\footnote {See ref.~\cite{deR01} for a
recent review, and refs.~\cite{KPdeR98} to \cite{KPdeR01} for
details.}, but which turn out to be useful as well for the
evaluation of this particular class of hadronic contributions to
$a_{\mu}$. This paper is dedicated to the evaluation of these
contributions.

\vskip 2pc \centerline{\epsfbox{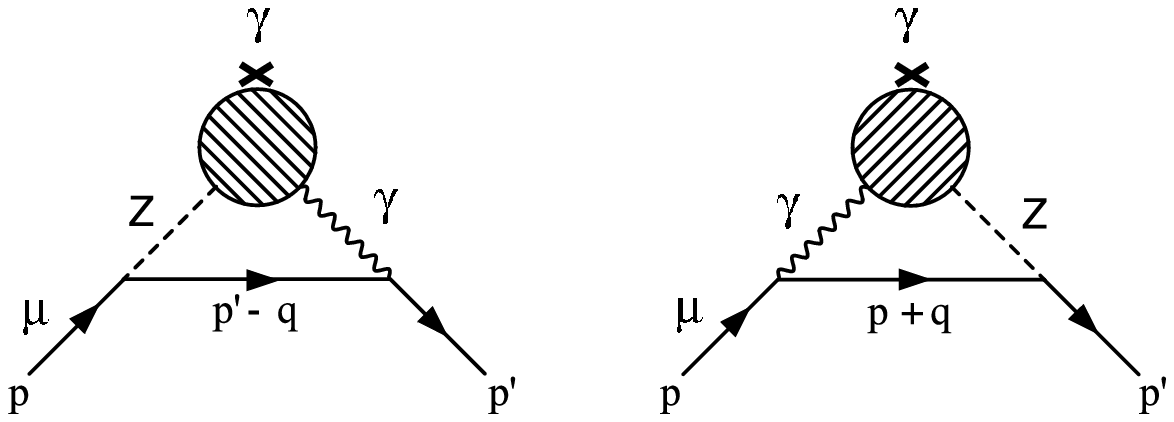}}\vspace*{0.5cm}
\centerline{{\bf Fig.~1} {\it Feynman diagrams with the hadronic
$\gamma-\gamma-Z$ vertex which contributes to the muon anomaly.}}
\vskip 2pc

The relevant terms in the Lagrangian of the Standard Theory which
we shall be needing are the following:
\begin{description}
\item[{\it 1a. The Leptonic Neutral Current Lagrangian}]
$$
\cL_{\mbox{\rm \tiny NC}}^{\mbox{\rm \tiny leptons}} = e
J_{\mu}^{\mbox{\rm\tiny em}}A^{\mu}-
\frac{g}{2\cos\theta_{W}}J_{\mu}^{(0)}Z^{\mu} \,,
$$
where
$$
J_{\mu}^{\mbox{\rm\tiny
em}}=\sum_{f}Q_{f}\bar{f}\gamma_{\mu}f\,,\quad\annd\quad
J_{\mu}^{(0)}=\sum_{f}\bar{f}\gamma_{\mu}(\mbox{\rm
v}_{f}-\mbox{\rm a}_{f}\gamma_{5})f\,,
$$
with
$$
\mbox{\rm v}_{f}=T_{3f}-2Q_{f}\sin^2\theta_{W}\,,\quad\annd\quad
\mbox{\rm a}_{f}=T_{3f}\,.
$$
In our case, we shall only need the muon components
$$
Q_{\mu}=-1\,,\quad\annd\quad\mbox{\rm a}_{\mu}=-1/2\,;
$$
i.e., the muon electromagnetic coupling and the axial coupling of the $Z$
to the muon line:
$$
-e\bar{\mu}(x)\gamma_{\alpha}\mu(x)A^{\alpha}(x)
\quad\annd\quad\frac{g}{2\cos\theta_{W}}\frac{1}{2}
\bar{\mu}(x)\gamma_{\alpha}\gamma_{5}\mu(x)Z^{\alpha}(x)\,.
$$

\item[{\it 1b. The Hadronic Neutral Current Lagrangian in the sector of
light quarks}]
$$
\cL_{\mbox{\rm \tiny NC}}^{\mbox{\rm \tiny
quarks}}=\bar{q}(x)\gamma^{\mu}\left[
l_{\mu}^{(0)}\gL+r_{\mu}^{(0)}\gR\right]q(x)\,,
$$
where $\bar{q}=(\bar{u},\bar{d},\bar{s})$ and
$$
l_{\mu}^{(0)}=eQ_{L}[A_{\mu}-\tan\theta_{W}Z_{\mu}]+
\frac{g}{2\cos\theta_{W}}Q_{L}^{(3)}Z_{\mu}\,\quad\annd\quad
r_{\mu}^{(0)}=eQ_{R}[A_{\mu}-\tan\theta_{W}Z_{\mu}]\,.
$$
In our calculation, only the hadronic electromagnetic
coupling:
$$
e\, \bar{q}(x)\gamma^{\mu}Qq(x)A_{\mu}\,, \qquad Q=Q_{L}=Q_{R}=
{\mbox{\rm \small diag}}\ (2/3,-1/3,-1/3)\,;
$$
and the hadronic axial coupling:
$$
-\frac{1}{2}\frac{g}{2\cos\theta_{W}}\bar{q}(x)\gamma^{\mu}\gamma_{5}
Q_{L}^{(3)}q(x)Z_{\mu}(x)\,,\qquad Q_{L}^{(3)}={\mbox{\rm \small diag}}\
(1,-1,-1)\,,
$$
will be needed.

\item[{\it 1c. The coupling of the unphysical neutral Higgs field to
the light quarks and to the muon}]
\be\lbl{unH1}
\cL_{\Phi^{0}}=\frac{g}{2\cos\theta_{W}}\Phi^{0}(x)J^{\Phi^{0}}(x)\,,
\ee
with (${\mbox{a}}_{u}=1/2\,,{\mbox{a}}_{d}={\mbox{a}}_{s}=-1/2$)
\be\lbl{unH2}
J^{\Phi^{0}}(x)=2\sum_{q=u,d,s}
{\mbox{a}}_{q}\frac{m_{q}}{M_{Z}}\bar{q}(x)i\gamma_{5}q(x)+
2\mbox{\rm a}_{\mu}\frac{m_{\mu}}{M_{Z}}\bar{\mu}(x)i\gamma_{5}\mu(x)\,.
\ee
\end{description}

\noi
The hadronic electroweak vertex which we have to
compute, in the notation corresponding to Fig.~1, is the
following~\footnote{We use the following conventions for Dirac's
$\gamma$--matrices:
$\{\gamma_{\mu},\gamma_{\nu}\}=2g_{\mu\nu}$, with $g_{\mu\nu}$ the
Minkowski metric tensor of signature $\{+,-,-,-\}$,
$\gamma_{5}=i\gamma^{0}\gamma^{1}\gamma^{2}\gamma^{3}$, and the totally
antisymmetric tensor  $\epsilon_{\mu\nu\rho\sigma}$ is such that
 $\epsilon_{0123}=+1\,.$}
$$
\langle\bar{\mu}(p')\vert
V_{\rho}^{\elm}(0)\vert\mu(p)\rangle=
\bar{u}(p')\Gamma_{\rho}(p',p)u(p)=(-ie)(-ie)
\left(\frac{ig}{4\cos\theta_{W}}\right)
\left(\frac{-ig}{4\cos\theta_{W}}\right)\times  \nn
$$
$$
\int\frac{d^4q}{(2\pi)^4}\,
\frac{-i}{q^2}\,\frac{-i}{(p'-p-q)^2-M_{Z}^2}
\bar{u}(p')
\left[\gamma^{\mu}\frac{i}{\pslsh'\ -\qsls-m_{\mu}}\gamma^{\nu}\gamma_{5}+
\gamma^{\nu}\gamma_{5}\frac{i}{\pslsh\ +\qsls-m_{\mu}}
\gamma^{\mu}\right]u(p)\times
$$
\be\lbl{vertex} \int d^4x\,e^{iq.x}\int d^4y\, e^{i(p'-p-q).y}
\langle \Omega\vert
T\{V_{\mu}^{\elm}(x)A_{\nu}^{\nc}(y)V_{\rho}^{\elm }(0)\}\vert
\Omega\rangle\,, \ee where
$$
V_{\mu}^{\elm}(x)=\bar{q}(x)\gamma_{\mu}\, Q\, q(x)\,,\qquad
A_{\nu}^{\nc}(y)=\bar{q}(y)\gamma_{\nu}\gamma_{5}\, Q_{L}^{(3)}\,
q(y)
$$
and $\vert\Omega\rangle$ denotes the full QCD physical vacuum.
Here, for the purpose of simplicity, both the photon and
$Z$ propagators in the second line of Eq.~\rf{vertex} are evaluated in the
Feynman gauge. We dedicate, however, Appendix A to a discussion
of questions related to gauge dependence. The anomalous magnetic moment
contribution from this vertex  is then defined by the corresponding Pauli
form factor at zero momentum transfer, i.e.,
\be\lbl{pauli}
F_{2}(0)=\lim_{k^2\ra
0}\tr\left\{(\pslsh+m_{\mu})\Lambda_{2}^{\rho}(p',p)(\pslsh'+m_{\mu})
\Gamma_{\rho}(p',p)
\right\}\,,
\ee
where $(p'=p+k)$
\be\lbl{projector}
\Lambda_{2}^{\rho}(p',p)=\frac{m_{\mu}^2}{k^2}\frac{1}
{4m_{\mu}^2-k^2}\gamma^{\rho}
-\frac{m_{\mu}}{k^2}\frac{2m_{\mu}^2+k^2}{(4m_{\mu}^2-k^2)^2}
(p+p')^{\rho}
\ee
is the projector on the Pauli form factor,
and $m_{\mu}$ denotes the muon mass.

\vspace{0.7cm}

\section{\normalsize The Master Green's Function}
\setcounter{equation}{0}
\lbl{gf}

\noi The contribution we want to compute is governed by the
hadronic Green's function \be\lbl{VAV} W_{\mu\nu\rho}(q,k)= \int
d^4x\,e^{iq\cdot x}\int d^4y\, e^{i(k-q)\cdot y} \langle
\Omega\,\vert
T\{V_{\mu}^{\elm}(x)A_{\nu}^{\nc}(y)V_{\rho}^{\elm}(0)\}\vert
\Omega\rangle\,, \ee with $k$ the incoming photon four--momentum
associated with the classical external magnetic field, as
illustrated in Fig.~2.

\vskip 2pc \centerline{\epsfbox{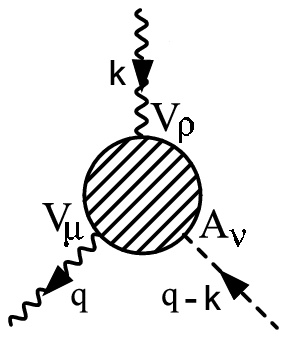}}\vspace*{0.5cm}
\centerline{{\bf Fig.~2} {\it Symbolic representation of the
$\langle V\!\!AV\rangle$ three--point function in Eq.~\rf{VAV}.}}
\vskip 2pc

\noi
The general Ward identities which
constrain this type of three--point functions in QCD are discussed in
Appendix A. In particular, $W_{\mu\nu\rho}(q,k)$ has an electroweak $U(1)$
anomalous term. It is this term which, when lumped together with
the corresponding contributions from the $c$ quark and the $e$
and $\mu$ leptons to the $\gamma-\gamma-Z$ loop, gives a finite
gauge invariant contribution to $a_{\mu}$ which was calculated,
independently, in refs.~\cite{PPdeR95,CKM95}.  Here we are
particularly interested in the contribution to $a_{\mu}$ from the
non--anomalous part of $W_{\mu\nu\rho}(q,k)$, denoted by
$\tilde{W}_{\mu\nu\rho}(q,k)$, i.e.
\be\lbl{tilde}
W_{\mu\nu\rho}(q,k)=W_{\mu\nu\rho}(q,k)_{\mbox{\rm\tiny
anomaly}}+\tilde{W}_{\mu\nu\rho}(q,k) \,, \ee where \be
W_{\mu\nu\rho}(q,k)_{\mbox{\rm\tiny anomaly}}=-
i\frac{N_c}{12\pi^2}\frac{4}{3}\frac{(q-k)_{\nu}}{(q-k)^2}
\epsilon_{\mu\rho\alpha\beta}q^{\alpha}k^{\beta}\,.
\ee
The function $\tilde{W}_{\mu\nu\rho}(q,k)$ obeys the
trivial Ward identities
\be\lbl{ward}
q^{\mu}\tilde{W}_{\mu\nu\rho}(q,k)=0\,,\quad\annd\quad
k^{\rho}\tilde{W}_{\mu\nu\rho}(q,k)=0\,,
\ee
but, as discussed in Appendix A, it has both a longitudinal and transverse
component in the axial neutral--current $\nu$--index:
$$
\tilde{W}_{\mu\nu\rho}(q,k)=\tilde{W}_{\mu\nu\rho}^{\mbox{\tiny
long}}(q,k) +\tilde{W}_{\mu\nu\rho}^{\mbox{\tiny trans}}(q,k)\,.
$$

Differentiating the
second identity in Eq.~\rf{ward} with respect to $k$  we get
\be\lbl{WQ2}
\tilde{W}_{\mu\nu\rho}(q,k)=-k^{\sigma} \frac{\partial}{\partial
k^{\rho}}\tilde{W}_{\mu\nu\sigma}(q,k)\,.
\ee
As we shall see in
the next section, extracting one power of $k$ from
$\tilde{W}_{\mu\nu\rho}^{\mbox{\tiny trans}}(q,k)$ is all that is needed
in order to obtain the corresponding contribution to the Pauli form factor
at zero momentum transfer, i.e. the anomalous magnetic moment $a_{\mu}$.
Therefore, without loss of generality, we can set
$$ \tilde{W}_{\mu\nu\rho}^{\mbox{\tiny trans}}(q,k)=
k^{\sigma} W_{\mu\nu\rho\sigma}(q)+\cO(k^2)\,,
$$
where
\be\lbl{Wk}
W_{\mu\nu\rho\sigma}(q)= -\frac{\partial}{\partial
k^{\rho}}\left[W_{\mu\nu\sigma}(q,k)-\frac{(q-k)_{\nu}}{(q-k)^2}
(q-k)^{\nu'}W_{\mu\nu'\sigma}(q,k)\right]\Big\vert_{k\ra
0}\,.
\ee
The most general pseudo--tensor $
W_{\mu\nu\rho\sigma}(q)$ which satisfies the Ward identities
is then constrained to have the form
$(Q^2\equiv -q^2)$~\footnote{There is {\it a priori} another possible
tensor structure which also fulfills these requirements:
$$
\epsilon_{\mu\nu\rho\sigma} q^2+
q_{\nu}\epsilon_{\mu\rho\lambda\sigma}q^{\lambda}+
q_{\mu}\epsilon_{\rho\nu\lambda\sigma}q^{\lambda}\,.
$$
However, using the identity
$$
\epsilon^{\mu\nu\rho\sigma}q^{\lambda}+
\epsilon^{\nu\rho\sigma\lambda}q^{\mu}+
\epsilon^{\rho\sigma\lambda\mu}q^{\nu}+
\epsilon^{\sigma\lambda\mu\nu}q^{\rho}+
\epsilon^{\lambda\mu\nu\rho}q^{\sigma}=0\,,
$$
it can easily be shown (contracting  the previous identity with
$q_{\lambda}$) that it is identical to the one in
Eq.~\rf{mastertilde}.}: \be\lbl{mastertilde}
W_{\mu\nu\rho\sigma}(q)=iW(Q^2)\left[q_{\rho}
\epsilon_{\mu\nu\alpha\sigma}q^{\alpha}-q_{\sigma}
\epsilon_{\mu\nu\alpha\rho}q^{\alpha}\right]\,. \ee As we shall
see in the next section, the contribution we are looking for can
be expressed as a simple integral of the function $W(Q^2)$ over
the range $0\le Q^2 \le\infty$ in the euclidean $Q^2$ variable.
We shall show in section~4 that, in the chiral limit,
this function falls as $1/Q^6$ at large $Q^2$ and, in the
presence of massive light quarks, the large--$Q^2$ fall--off of
$W(Q^2)$ goes as $1/Q^4$ only.

\vspace{0.7cm}

\section{\normalsize The Contribution to the Muon Anomalous Magnetic
Moment}
\setcounter{equation}{0}
\lbl{integral}

\noi
Since what we want is $F_{2}(0)$ we need only to keep the
projector in Eq.~\rf{projector} to its simplest form in an
expansion in powers of momentum $k$,
$$
\Lambda_{2}^{\rho}(p',p)=\frac{1}{4k^2}\left(\gamma^{\rho}
-\frac{(p+p')^{\rho}}{2m_{\mu}}\right) + \cdots\,.
$$
This is because
the product
$$
(\pslsh+m_{\mu})\Lambda_{2}^{\rho}(p',p)(\pslsh'+m_{\mu})=
\frac{1}{4k^2}(\pslsh+m_{\mu})
\left[\gamma^{\rho}\ksls-\left(k^{\rho}+
\frac{p^{\rho}}{m_{\mu}}\ksls\right)\right]+\cO(k^0)\,,
$$
which
then appears in Eq.~\rf{pauli}, is proportional to one inverse
power of $k\,$. On the other hand, gauge invariance
forces the Green's function $W_{\mu\nu\rho}(q,k)$ to be
proportional to one power of $k$ at least. Altogether, this gives
the required powers of $k$ which are needed to define
$F_{2}(0)$. We can, therefore, simplify significantly the
calculation of Eq.~\rf{pauli} to that of the simpler
expression:
$$
F_{2}(0)=(-e^2)\frac{g^2}{16\cos^2\theta_{W}}
\frac{1}{M_{Z}^2}
\lim_{k^2\ra
0}\int\frac{d^4q}{(2\pi)^4}\frac{1}{q^2}
\left(\frac{M_{Z}^2}{q^2 -M_{Z}^2}\right)\times
$$
$$
\frac{1}{4k^2}\tr\left\{(\pslsh+m_{\mu})
\left[\gamma^{\rho}\ksls-\left(k^{\rho}+
\frac{p^{\rho}}{m_{\mu}}\ksls\right)\right]
\left[\gamma^{\mu}\frac{(\pslsh\ -\qsls+m_{\mu})}
{q^2-2q\dd p}
\gamma^{\nu}\gamma_{5}+\gamma^{\nu}\gamma_{5}
\frac{(\pslsh\ +\qsls+m_{\mu})}{q^2+2q\dd p}\gamma^{\mu}
\right]
\right\}\times
$$
\be\lbl{ano}
\left[-\frac{N_c}{12\pi^2}\frac{4}{3}
\frac{q_{\nu}}{q^2}\epsilon_{\mu\rho\alpha\sigma}q^{\alpha}k^{\sigma}
-i\tilde{W}_{\mu\nu\rho}^{\mbox{\rm\tiny long  }}(q,k)+
k^{\sigma}\left[q_{\rho}
\epsilon_{\mu\nu\alpha\sigma}q^{\alpha}-q_{\sigma}
\epsilon_{\mu\nu\alpha\rho}q^{\alpha}\right]W(Q^2)\right]\,,
\ee
where the first term in the last line of Eq.~\rf{ano} is the
contribution from the anomaly to the $\langle V\!\!AV\rangle$ Green's
function, already discussed in
ref.~\cite{PPdeR95}. The second and third terms are the non--anomalous
contributions from the longitudinal and transverse components of the
same Green's function. Let us first simplify, still further, the last
contribution.

The integral over the four--momentum
$q$ from the last term (the transverse component), is convergent in the
infrared because
$W(Q^2)$ has no Goldstone pole at $Q^2\ra 0$ and in the ultraviolet
because of the QCD short--distance behaviour of $W(Q^2)$, which we shall
later discuss. Therefore, to leading
order in powers of the lepton mass, we can perform the trace and then
the integration over the angles, setting
\be\lbl{app}
\frac{1}{q^2\pm 2q\dd p}\ra \frac{1}{q^2}\,,
\ee
in Eq.~\rf{ano}. The integral over the solid angle
$d\Omega_{q}$ becomes trivial, and the transverse contribution
is then given by the expression
$$
F_{2}(0)\big\vert_{\mbox{\rm\tiny
trans}}=\frac{g^2}{16\cos^2\theta_{W}}\frac{e^2}
{M_{Z}^2}\lim_{k^2\ra 0}\frac{i}{16\pi^2}\int_{0}^{\infty}dQ^2
Q^2\frac{1}{Q^4}
\left(\frac{M_{Z}^2}{Q^2+M_{Z}^2}\right)Q^2
W(Q^2)\times
$$
$$
\frac{-1}{8k^2}\tr\left\{(\pslsh+m_{\mu})
\left[\gamma^{\rho}\ksls-\left(k^{\rho}+
\frac{p^{\rho}}{m_{\mu}}\ksls\right)\right]\left[\gamma^{\mu}
(\pslsh+m_{\mu})
\gamma^{\nu}\gamma_{5}+\gamma^{\nu}\gamma_{5}
(\pslsh+m_{\mu})\gamma^{\mu}
\right]
\right\}\  k^{\sigma}\epsilon_{\mu\nu\rho\sigma}
$$
\be\lbl{g-2}
+\cO\left(\frac{m_{\mu}^2}{M_{Z}^2}\log\frac{M_{Z}^2}{m_{\mu}^2}
\right)\,.
\ee
The remaining algebra leads to a remarkably simple
expression:
\be\lbl{intW}
F_{2}(0)\big\vert_{\mbox{\rm\tiny
trans}}=(-e^2)\frac{g^2}{16\cos^2\theta_{W}}\left(\frac{m_{\mu}^2}
{M_{Z}^2}\right)\frac{1}{4\pi^2}\int_{0}^{\infty}dQ^2
\left(\frac{M_{Z}^2}{Q^2+M_{Z}^2}\right)W(Q^2)\,.
\ee

The integration of the anomalous term in Eq.~\rf{ano} is in fact
slightly more complicated, because there we cannot simplify the muon
propagators as in Eq.~\rf{app}. It is then useful to combine
denominators using two Feynman parameters:
$$
\frac{1}{(q^2)^2 (q^2-M_{Z}^2)(q^2\pm 2q\cdot p)}=6\int_{0}^{1}dx x^2
\int_{0}^{1}dy y\frac{1}{\left[(q-l)^2-R^2 \right]^4}\,,
$$
where
$$
-l=\pm p(1-x)\qquad\annd\qquad R^2=M_{Z}^2\
x(1-y)+m_{\mu}^{2}(1-x)^2\,.
$$
Then, after the shift $q\ra q+l$ of the integration
four--momentum, and upon neglecting terms of
$\cO\left(\frac{m_{\mu}^2}{M_{Z}^2}\log\frac{M_{Z}^2}
{m_{\mu}^2}\right)$ one obtains the result (in the Feynman gauge)

{\setl
\bea
F_{2}(0)\big\vert_{\mbox{\rm\tiny
anom}} & = & \frac{\GF}{\sqrt{2}}\frac{m_{\mu}^2}{8\pi^2}
\frac{\alpha}{\pi}\frac{N_c}{3}\frac{4}{3}
\times  \nn \\
& & \int_{0}^{1}\!dx
\,x^2
\int_{0}^{1}\!dy\,y\!\left[
\frac{1+3(1-x)}{x(1-y)+\frac{m_{\mu}^2}{M_{Z}^2}(1-x)^2}
-\frac{m_{\mu}^2}{M_{Z}^2}
\frac{(1-x)^3}{\left[x(1-y)+\frac{m_{\mu}^2}{M_{Z}^2}(1-x)^2
 \right]^2}\right]\,, \nn
\\\lbl{anomal} & = & \frac{\GF}{\sqrt{2}}\frac{m_{\mu}^2}{8\pi^2}
\frac{\alpha}{\pi}N_c\left\{\frac{4}{9}
\log\frac{M_{Z}^2}{m_{\mu}^2}+\frac{2}{9}+\cO\left(\frac{m_{\mu}^2}
{M_{Z}^2}\log\frac{M_{Z}^2}
{m_{\mu}^2}\right)\right\}\,, \nn \\
 &= & \frac{\GF}{\sqrt{2}}\frac{m_{\mu}^2}{8\pi^2}
\frac{\alpha}{\pi}\times 18.69\,,
\eea}

\noi
in agreement with the result~\footnote{The constant term inside  the
bracket in Eq.~(20) of ref.~\cite{PPdeR95} should be $2/3$ instead of
$4/9$.} in ref.~\cite{PPdeR95}. The evaluation of
$F_{2}(0)$ in an arbitrary gauge, with a discussion of the physics behind
it, can be found in Appendix A.

\vspace{0.7cm}

\section{\normalsize Constraints on the Function $W(Q^2)$ from the
Operator Product Expansion (OPE)}
\setcounter{equation}{0}
\lbl{Wfunction}

\noi
In order to perform the integral over $Q^2$ in Eq.~\rf{intW} we require
information on the function $W(Q^2)$. We shall now discuss the
constraints on this function, at large $Q^2$ values, which follow from
QCD at short--distances.
For pedagogical reasons, we shall first discuss the
hadronic contributions  in the chiral limit, where
the light quark masses are neglected. However, the leading effect due
to the explicit chiral symmetry breaking induced by the finite light
quark masses, turns out to be important and rather subtle. It is at the
origin of a conceptual discrepancy with the treatment reported in
ref.~\cite{CKM95}. We shall, therefore, discuss as well the contributions
from explicit breaking, later, in section 6.

Since one of the photons is on shell, the relevant
short--distance behaviour we are looking for is the one given by
the OPE of the two currents $V_{\mu}^{\elm}(x)$ and
$A_{\nu}^{\nc}(y)$ in the three--point function
$\tilde{W}_{\mu\nu\rho}(q,k)$ in Eq.~\rf{VAV} when $x\ra y$,
because these are the currents where the hard virtual momentum
$q$ can get through \cite{SV} ( from here onwards, the subscript
{\it non--anomalous} in $T$--products will be understood). This
three--point function is represented symbolically in Fig.~2. The
quantity we are interested in is defined by the limit
\be\lbl{Ufunction} \cU_{\mu\nu}(q)=\lim_{q\ra\infty}\int d^4z
e^{iq.z} T\left\{V_{\mu}^{\elm}(z)A_{\nu}^{\nc}(0)\right\}\,, \ee
and the relevant term in the OPE of these two currents is the one
with a tensor structure \be\lbl{opeterm}
\cU_{\mu\nu}(q)=i\left[q_{\delta}
\epsilon_{\mu\nu\alpha\beta}q^{\alpha}-
q_{\beta}\epsilon_{\mu\nu\alpha\delta}q^{\alpha}\right]
\cC\frac{\cO^{\beta\delta}(0)}{\left(Q^2 \right)^p}+\cdots\,, \ee
where $\cC$ is a dimensionless constant to be calculated,  $p$
denotes the lowest possible power and the ellipsis stands for
other contributions proportional to higher powers of $1/Q^2$
and/or to other tensor structures which cannot contribute to the
muon anomaly. The dimension of the $\cO^{\beta\delta}$ operator
is such that:
$$
d\left[\cO^{\beta\delta}\right]-2p=0\,;
$$
i.e., $\cO^{\beta\delta}$ must be an operator of {\it even} dimension,
and antisymmetric in $\beta\delta$.

There are in fact two
types of potential contributions to the r.h.s. of Eq.~\rf{opeterm}
which, physics--wise, are generated by the emission of {\it
soft virtual gluons} and {\it hard virtual gluons} in the
$T$--product in Eq.~\rf{Ufunction}. As discussed in Appendix B, there is
no possible contribution from {\it soft gluons}, on symmetry grounds,
with the tensor structure of Eq.~\rf{opeterm} and with a power $p\le 3$.
The leading contribution, which appears with a power $p=3$, comes
from the perturbation theory expansion in {\it hard virtual
gluons} and it is the one shown in the diagrams of Fig.~3 below.

\vskip 2pc \centerline{\epsfbox{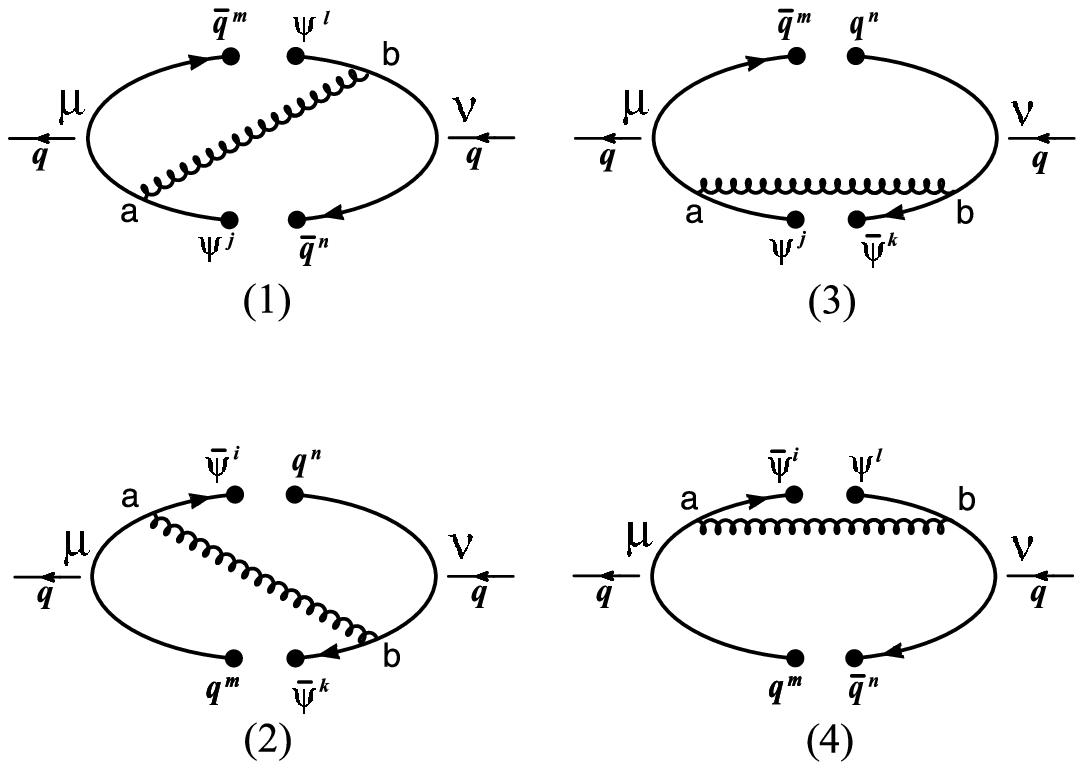}}\vspace*{0.5cm}
\noi {\bf Fig.~3} {\it Feynman diagrams showing the hard gluon
exchanges which contribute  to the leading $1/Q^6$ behaviour of the
operator $\cU_{\mu\nu}(q)$ in Eq.~\rf{Ufunction} .}
\vskip 2pc

\noi
The details of the calculation of the
contribution from {\it hard virtual gluons} to $\cU_{\mu\nu}(q)$ in
Eq.~\rf{Ufunction} are explained in Appendix B where it is shown that,
\be\lbl{operesult}
\cU_{\mu\nu}(q)=i\left[q_{\delta}
\epsilon_{\mu\nu\alpha\beta}q^{\alpha}-
q_{\beta}\epsilon_{\mu\nu\alpha\delta}q^{\alpha}\right]
\left(-2\pi^2\frac{\als}{\pi}\right)\frac{\cO^{\beta\delta}(0)}{\left(Q^2
\right)^3}+\cdots\,,
\ee
and where the operator $\cO^{\beta\delta}(0)$ is found to be the
four--quark operator:
$$
\cO^{\beta\delta}(0)=\left[\frac{2}{3}\left( \bar{u}\sigma^{\beta\delta}
u\right)\left(\bar{u}u \right)+\frac{1}{3}\left(
\bar{d}\sigma^{\beta\delta} d\right)\left(\bar{d}d
\right)+\frac{1}{3}\left( \bar{s}\sigma^{\beta\delta}
s\right)\left(\bar{s}s \right)\right](0)\,.
$$
We find, therefore, that the constant $\cC$ in Eq~\rf{opeterm},
to lowest
order in perturbative QCD is
$$
\cC= -2\pi^2\frac{\als}{\pi}+\cO\left(\als^2\right)\,.
$$

We are now in the position to evaluate the high--$Q^2$ behaviour of
the invariant function $W(Q^2)$ in Eq.~\rf{mastertilde}. For that we
first insert the result obtained  in Eq.~\rf{operesult} into the r.h.s.
of Eq.~\rf{Wk}. What appears then is the two--point function
$$
\Psi^{\beta\delta}_{~~\rho}(k)=\int d^4 y e^{-ik\cdot y}\langle
\Omega\,\vert  T\left\{\cO^{\beta\delta}(0) V_{\rho}^{\elm}
(y)\right\}\vert \Omega\rangle\,.
$$
In our case, the four--momentum $k$ in this two--point function is a {\it
soft momentum}; and in fact, what we need for our purposes is only the
term linear in this soft momentum, i.e.
{\setl
\bea
-\frac{\partial}{\partial
k^{\rho}}\Psi^{\beta\delta}_{~~\sigma}(k)\Big\vert_{k\ra 0} & = & i
\int d^4 y\ y_{\rho}\ \langle \Omega\,\vert
T\left\{\cO^{\beta\delta}(0) V_{\sigma}^{\elm}
(y)\right\}\vert \Omega\rangle \lbl{VT1} \\
& = & i\stern_{0}
\int d^4 y\ y_{\rho}\ \langle \Omega\,\vert
T\left\{ V_{\sigma}^{\elm }
(y)\ \left(\bar{q}(0)\sigma^{\beta\delta} Q Q_{L}^{(3)}
q(0)\right)\right\}\vert \Omega\rangle\,, \lbl{VT2}
\eea}

\noi
where in going from Eq.~\rf{VT1} to Eq.~\rf{VT2} we have used the
$1/N_c$--expansion, keeping only the leading contribution, and $SU(3)$
symmetry for the single flavour vacuum  condensate $\stern_{0}$. Our
problem is then reduced to the one of evaluating the behaviour of the
$\langle VT\rangle$ correlation function which appears in the r.h.s. of
Eq.~\rf{VT2} at a small momentum transfer.

On general grounds (conservation of the vector current, parity
invariance and $SU(3)$ symmetry) and in the chiral limit, the generic
$\langle VT\rangle$ correlation function
\be\lbl{VTfunc}
\int d^4y\ e^{ik\cdot y}\langle \Omega\,\vert
T\left\{\left(\bar{\psi}\gamma_{\sigma}\frac{\lambda^{a}}{2}\psi\right)
(y)\ \left(\bar{\psi}\sigma^{\beta\delta}\frac{\lambda^{b}}{2}
\psi\right)(0)\right\}\vert \Omega\rangle = (k^{\beta}
\delta_{\sigma}^{\delta}- k^{\delta}\delta_{\sigma}^{\beta})\
\delta^{ab}\
\Pi_{VT}(k^2)\,,
\ee
depends on one invariant function $\Pi_{VT}(k^2)\,,$  where
$\lambda^a$ denotes the Gell-Mann matrices,
normalized as tr$(\lambda^a\lambda^b)=2\delta^{ab}$, with
$\lambda^0=(\sqrt{2/3})\,\mbox{\bf 1}\,,$ and
$a=0,1,\dots\,8$. Using the
relations
$$
Q = \frac{\lambda^3}{2} + \frac{1}{\sqrt
3}\frac{\lambda^8}{2}\,,
\quad\annd\quad
QQ_{L}^{(3)} =
\frac{1}{3}\Big(\frac{\lambda^3}{2} +\frac{1}{\sqrt
3}\frac{\lambda^8}{2}+4\sqrt{\frac{2}{3}}\frac{\lambda^0}{2}\Big)\,,
$$
we can write our result in Eq.~\rf{VT2} in terms of $\Pi_{VT}(0)$
only, as follows:
$$
-\frac{\partial}{\partial
k^{\rho}}\Psi^{\beta\delta}_{~~\sigma}(k)\Big\vert_{k\ra
0}=\frac{1}{3}\left(1+\frac{1}{3}\right)\stern_{0}
(\delta^{\beta}_{\rho} \delta_{\sigma}^{\delta}-
\delta^{\delta}_{\rho}\delta_{\sigma}^{\beta})\Pi_{VT}(0)\,.
$$

The function
$\Pi_{VT}(k^2)\,,$ in the {\it minimal hadronic approximation} (MHA) to
large--$N_c$ QCD, has been evaluated in ref.~\cite{KN01}. In this
approximation, it has the simple pole form
$$
\Pi_{VT}(k^2)\big\vert_{\mbox{\rm\tiny
MHA}}=-\stern_{0}\frac{1}{k^2-M_{\rho}^2}\,,
$$
with $M_{\rho}$ the mass of the lowest vector state in the large--$N_c$
QCD spectrum. The asymptotic behaviour of the function  $W(Q^2)$ at
large $Q^2$ values is then fixed, in the MHA, with the result
\be\lbl{opefall}
\lim_{Q^2\ra\infty}W(Q^2)\big\vert_{\mbox{\rm\tiny MHA
}}=\frac{16}{9}\pi^2\frac{\als}
{\pi}\frac{1}{M_{\rho}^2}\frac{\stern_{0}^2}{Q^6}\,.
\ee

\vspace{0.7cm}

\section{\normalsize The Function $W(Q^2)$ in the MHA to Large--$N_c$
QCD and in the Chiral Limit}
\setcounter{equation}{0}

\noi Since $W(Q^2)$ has dimensions of an inverse squared mass, it
is convenient to define \be\lbl{func}
W(Q^2)=\frac{1}{M_{\rho}^2}w(z)\,,\quad\with\quad
z\equiv\frac{Q^2}{M_{\rho}^2}\,. \ee The function $w(z)$ in
large--$N_c$ QCD is a meromorphic function in the complex
$z$--variable. It has the pole structure shown by the Feynman
diagrams in Fig.~4, where the {\Large$\rho_{i}$} denote vector
narrow states and the {\Large a}$_{j}$ axial--vector states.

\vskip 2pc \centerline{\epsfbox{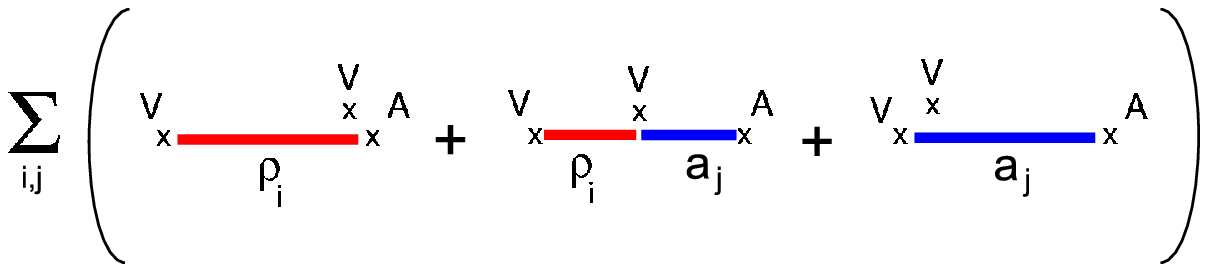}}\vspace*{0.5cm} \noi
\centerline{{\bf Fig.~4} {\it Feynman diagrams which contribute to
the function $w(z)$ in Eq.~\rf{func} in large--$N_c$ QCD.}} \vskip
2pc

\noi
Using partial fractions, we can write in full generality
$$
w(z)=\sum_{R}\frac{\alpha_{R}}{z+z_{R}}\,,\quad\with\quad
z_{R}=\frac{M_{R}^2}{M_{\rho}^2}\,,
$$
and $M_{R}$ the mass of the $R$ narrow state.
The short--distance constraints on the unknown residues $\alpha_{R}$ are
$$
\sum_{R}\alpha_{R}  =  0\,,\quad
\sum_{R}\alpha_{R}z_{R} =  0 \,,\quad\annd\quad
\sum_{R} \alpha_{R}z_{R}^2  =  \cR \,,
$$
with $\cR$ fixed by the leading term in the OPE expansion
$$
\lim_{z\ra\infty} w(z)=\cR\frac{1}{z^3}+ \cO(\frac{1}{z^4})\,.
$$

\noi
Therefore, in large--$N_c$ QCD
\be\lbl{intN}
F_{2}(0)\big\vert_{\mbox{\rm\tiny
trans}}=(-e^2)\frac{g^2}{16\cos^2 \theta_{W}}\left(\frac{m_{\mu}^2}{M_Z^2}
\right)\frac{1}{4\pi^2}\int_{0}^{\infty} dz
\frac{1}{1+\frac{M_{\rho}^2}{M_{Z}^2}z}\sum_{R}
\frac{\alpha_{R}}{z+z_{R}}\,.
\ee

The minimum number $p$ of poles required to have the same
asymptotic behaviour as the OPE result in Eq.~\rf{opefall} is
$p=3$. This minimum number of poles corresponds in this case to
the {\it minimal hadronic approximation} to large--$N_c$ QCD,
which we have used already in several other
calculations~\footnote{See ref.~\cite{deR01} for a recent review,
and refs.~\cite{KPdeR98} to \cite{KPdeR01} for details.}. In this
approximation, the expansion of the meromorphic function $w(z)$ in
rational functions~\footnote{Recall the Mittag--Leffler theorem
for meromorphic functions, see e.g. ref.~\cite{WW} section {\bf
7.4} } is limited to the first {\it three poles}, and can be
written as follows: \be\lbl{mha3poles}
w(z)=\frac{\alpha_{1}}{z+1}+
\frac{\beta_{1}}{z+\frac{1}{g_{A}}}+\frac{\alpha_{2}}{z+r'}\,, \ee
with $\alpha_{1}$, $\beta_{1}$ and $\alpha_{2}$ unknown residues,
which we shall determine readily, and
$$
g_{A}=M_{\rho}^2/M_{A_{1}}^2\quad\annd\quad r'=M_{\rho'}^2/M_{\rho}^2\,.
$$
As we have seen, the OPE evaluated in the MHA, fixes the value of $\cR$
to
\be\lbl{Rope}
\cR=\frac{16}{9}\pi^2\frac{\als}
{\pi}\frac{\stern_{0}^2}{M_{\rho}^6}\,.
\ee
There
follow then  three short--distance constraints on the three
unknown residues
$\alpha_{1}$,
$\alpha_{2}$ and $\beta_{1}$:
{\setl
\bea
\alpha_{1}+\beta_{1}+\alpha_{2} & = &0\,, \nn
\\
\alpha_{1}+\frac{1}{g_{A}}\beta_{1}+r'\alpha_{2} & = & 0\,, \nn \\
\alpha_{1}+\frac{1}{g_{A}^2}\beta_{1}+r'^{2}\alpha_{2} & = & \cR \nn
\,,
\eea}

\noi
with the solution:
{\setl
\bea
\alpha_{1} & = & \cR\frac{g_{A}}{(1-g_{A})(r'-1)}\,, \nn
 \\
\alpha_{2} & = & -\cR\frac{g_{A}}{(1-g_{A}r')(r'-1)}\,, \nn \\
\beta_{1} & = & \cR\frac{g_{A}^2}{(1-g_{A})(1-g_{A}r')}\,. \nn
\eea}

\noi
In particular
$$
w(0)=\alpha_{1}+g_{A}\beta_{1}+\frac{1}{r'}\alpha_{2}=
\cR\frac{g_{A}}{r'}\,.
$$
The function $w(z)$ is then fully determined and the integral
which defines $F_{2}(0)$ in Eq.~\rf{g-2} can be done. The shape
of the function $w(z)$ normalized to its value at the origin is
the continuous red curve shown in Fig.~5 below. As expected, it
is a sharply decreasing function in $z$.

At this stage, it is perhaps useful to compare the function
$W(Q^2)$ we find in the MHA to large--$N_c$ QCD with the
corresponding function, evaluated in the constituent chiral quark
model ($\chi$QM), which was used in ref.~\cite{PPdeR95} to obtain
$F_{2}(0)\big\vert_{\mbox{\rm\tiny trans}}$ as an estimate of the
error in $F_{2}(0)$. The function $W(Q^2)$ in the  $\chi$QM has a
simple parametric form: \be\lbl{param}
W_{\mbox{$\chi$\rm\footnotesize
QM}}(Q^2)=\frac{N_c}{12\pi^2}\frac{8}{3} g_{A}\ \int_{0}^{1}dx
\int_{0}^{1-x}dy \frac{xy-y(1-y)}{M_{Q}^2+Q^2 y(1-y)}\,, \ee
where $g_{A}\simeq 1/2$ is the axial--coupling of the constituent
quark\footnote{See the second reference in \cite{BBdeR93}. }.
With the same normalization as in Eq.~\rf{func}, i.e.,
\be\lbl{qmfunc} W_{\mbox{$\chi$\rm\footnotesize
QM}}(Q^2)=\frac{1}{M_{\rho}^2}w_{\mbox{$\chi$\rm\footnotesize
QM}}(z)\,,\quad\annd\quad z\equiv\frac{Q^2}{M_{\rho}^2}\,, \ee we
have

{\setl \bea\lbl{WchiQM} w_{\mbox{$\chi$\rm\footnotesize QM}}(z) &
= & \frac{N_c}{12\pi^2}\frac{8}{3} g_{A}\frac{M_{\rho}^2}{M_{Q}^2}
\times \int_{0}^{1}dx \int_{0}^{1-x}dy
\frac{xy-y(1-y)}{1+\frac{Q^2}{M_{Q}^2} y(1-y)} \nn \\
 & = & \lbl{qmw} \frac{-N_c}{12\pi^2}\frac{8}{3}
g_{A}\frac{M_{\rho}^2}{4Q^2}\left\{ 1+\frac{M_{Q}^2}{Q^2}
\frac{2}{\sqrt{1+\frac{4M_{Q}^2}{Q^2}}}
\log{\frac{\sqrt{1+\frac{4M_{Q}^2}{Q^2}}-1}
{\sqrt{1+\frac{4M_{Q}^2}{Q^2}}+1}}\right\}\,. \eea}

\noi The asymptotic behaviours of this function are:
$$
\lim_{z\ra 0}w_{\mbox{$\chi$\rm\footnotesize
QM}}(z)=\frac{-N_c}{12\pi^2}
g_{A}\frac{M_{\rho}^2}{9M_{Q}^2}+\cO\left(\frac{Q^2}{M_{Q}^2}\right)\,,
$$
and
$$
\lim_{z\ra \infty}w_{\mbox{$\chi$\rm\footnotesize
QM}}(z)=\frac{-N_c}{12\pi^2}\frac{2}{3}
g_{A}\frac{M_{\rho}^2}{Q^2}+
\cO\left[\left(\frac{M_{Q}^2}{Q^2}\right)^2\log\frac{Q^2}{M_{Q}^2}
\right]\,.
$$
The $\chi$QM not only fails to reproduce the large $Q^2$
behaviour of the OPE in QCD, but it has the opposite sign to the
MHA to large--$N_c$ QCD in the chiral limit. Notice
that, even though one is interested in the large-$Q^2$ limit,
there is one photon whose momentum $k^2$ vanishes (see Fig.~2).
This makes the Green's function $W(Q^2)$ in Eq. \rf{mastertilde} a
nonperturbative object which cannot be calculated in terms of
free quarks, such as those of the $\chi$QM.

The shape of the function $w_{\mbox{$\chi$\rm\footnotesize
QM}}(z)$ normalized to its value at the origin is the dashed curve
plotted in Fig.~5. It is yet another example of a constituent
chiral quark model prediction which deviates substantially from
the short--distance behaviour of QCD. The resulting contribution
to $F_{2}(0)\big\vert_{\mbox{\rm\tiny trans}}$ in the $\chi$QM is
as follows:

{\setl
\bea\lbl{CchiQM}
F_{2}(0)\Big\vert_{\mbox{\rm\tiny
trans}}^{\mbox{\rm\tiny $\chi\!\!$
QM}} & = &\frac{\GF}{\sqrt{2}}\
\frac{m_{\mu}^2}{8\pi^2}\ \frac{\alpha}{\pi}
\times \frac{N_c}{3}g_{A}\left[\frac{2}{3}
\log\frac{M_{Z}^2}{M_{Q}^2}-\frac{4}{3}+
\cO\left(\frac{M_{Q}^2}{M_{Z}^2}\log\frac{M_{Z}^2}{M_{Q}^2}\right)
\right]\,, \nn \\
 & = & \frac{\GF}{\sqrt{2}}\
\frac{m_{\mu}^2}{8\pi^2}\ \frac{\alpha}{\pi}
\times  3.08\,,\quad\foor\quad M_{Q}=330\,\MeV\quad\annd\quad
g_{A}=1/2\,.
\eea}

\vskip 2pc \centerline{\epsfbox{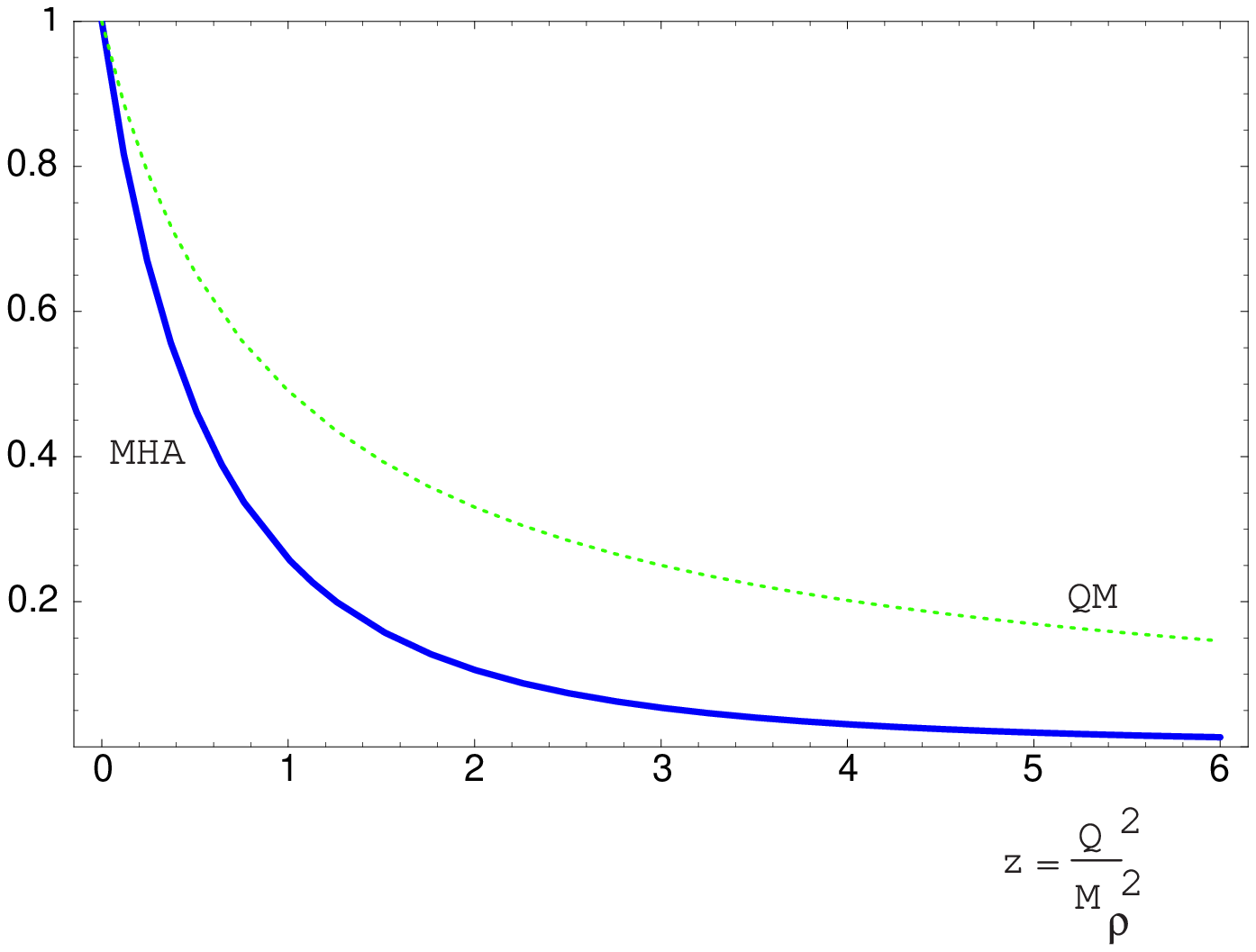}} \vspace*{1cm} {\bf
Fig.~5} {\it Shape of the functions $w(z)$ in Eq.~\rf{func} (the
continuous curve) and $w_{\chi QM}(z)$ in Eq.~\rf{qmfunc} (the
dotted  curve) normalized to their respective values at the
origin.} \vskip 2pc

Amusingly, the integral in
Eq.~\rf{intN}, in the MHA to
large--$N_c$ QCD can be easily done analytically. Upon neglecting terms of
$\cO\left(\frac{M_{\rho'}^2}{M_{Z}^2}\right)$, the result reads
$$
F_{2}(0)\big\vert_{\mbox{\rm\tiny
trans}}^{\mbox{\rm\tiny MHA}}=-\frac{\GF m_{\mu}^2}{\sqrt{2}}
\frac{\alpha}{2\pi}\cR
\left[\frac{g_{A}}{(1-g_{A}r')(r'-1)}\log{r'}+
\frac{g_{A}^2}{(1-g_{A})(1-g_{A}r')}\log{g_{A}}
\right]\,.
$$
Numerically, for $g_{A}=1/2$ and
$r'=3.5$, we find
\be\lbl{final}
F_{2}(0)\big\vert_{\mbox{\rm\tiny
trans}}^{\mbox{\rm\tiny MHA}}=-\frac{\GF m_{\mu}^2}{\sqrt{2}}
\frac{\alpha}{2\pi}\cR\times 0.128\,.
\ee

\noi
In order to evaluate, in terms of hadronic parameters, the
short--distance residue
$\cR$ in Eq.~\rf{Rope}, we observe that the same factor $\sim \frac{\als}
{\pi}\stern_{0}^2$ appears in the OPE of the $\langle
LR\rangle$ correlation function
where it has been shown~\cite{KdeR98} that in large--$N_c$ QCD
$$
 \frac{\als}
{4\pi}\stern_{0}^2 =\frac{1}{16\pi^2}
\left( \sum_{j}f_{A_{j}}^2 M_{A_{j}}^6
-\sum_{i}f_{V_{i}}^2 M_{V_{i}}^6\right)\,.
$$
In this case, the MHA corresponds to a saturation with a vector state and
an axial--vector state only, with the result
\be\lbl{MHALR}
 \frac{\als}
{4\pi}\stern_{0}^2\Big\vert_{\mbox{\rm\tiny MHA}}
=\frac{1}{16\pi^2}
\left(f_{A_{1}}^2 M_{A_{1}}^6
-f_{\rho}^2 M_{\rho}^6\right)\,.
\ee
Using this relation, as well
as the familiar vector dominance expressions~\cite{EGLPdeR89}:
$$
f_{\rho}^2=2\frac{F_{0}^2}{M_{\rho}^2}\,,\qquad f_{A_{1}}^2=
\frac{F_{0}^2} {M_{A_{1}}^2}\,,\qquad\annd\qquad
M_{\rho}^2=g_{A}\ M_{A_{1}}^2\,,
$$
which as explained in ref.~\cite{deR01} can also be viewed as predictions
of the MHA  to large--$N_c$  QCD,  our hadronic estimate
for the residue
$\cR$ is then
$$
\cR=\frac{8}{9}\frac{F_{0}^2}{M_{\rho}^2}\,,
$$
with $F_{0}$ the pion coupling constant in the chiral limit
($F_0=87~\MeV$). This estimate allows for a neat comparison with the
contributions calculated in refs.~\cite{PPdeR95,CKM95}, since we now have

{\setl
\bea\lbl{intMHA}
F_{2}(0)\big\vert_{\mbox{\rm\tiny
trans}}^{\mbox{\rm\tiny MHA}} & = & -\frac{\GF
}{\sqrt{2}}\frac{m_{\mu}^2}{8\pi^2}
\frac{\alpha}{\pi}\times
\left(\frac{2}{9}\times 0.128\right)\frac{16\pi^2 F_{0}^2}{M_{\rho}^2}\nn
\\
 & = & -\frac{\GF }{\sqrt{2}}\frac{m_{\mu}^2}{8\pi^2}
\frac{\alpha}{\pi}\times 0.06\,,
\eea}

\noi
Even allowing for a substantial error in our determination of
$\stern_{0}$ in Eq.~\rf{MHALR},  we obtain, in the chiral limit, a much
smaller contribution than the one predicted by the $\chi$QM in
Eq.~\rf{CchiQM}.

\vspace{0.7cm}

\section{\normalsize Beyond the Chiral Limit}
\setcounter{equation}{0}
\lbl{bcl}

\noi
As explained in
Appendix A, the axial Ward identity in Eq.~\rf{axialwiA}, in the presence
of light quark mass terms in the QCD Lagrangian, is no longer given by the
anomaly alone. As a result, the non--anomalous part of the Green's
function
$W_{\mu\nu\rho}(q,k)$ in Eq.~\rf{VAV}, which we have been
denoting by $\tilde{W}_{\mu\nu\rho}(q,k)$ (see Eq.~\rf{tilde}) is no
longer fully transverse. It acquires a longitudinal component
\be\lbl{fullwi}
\tilde{W}_{\mu\nu\rho}^{\mbox{\rm\tiny long }}(q,k)=-2iM_{Z}
\frac{(q-k)_{\nu}}{(q-k)^2}W_{\mu\rho}^{(\Phi^{0})}
(q,k)\,
\ee
where

{\setl
\bea\lbl{fullWI}
W_{\mu\rho}^{(\Phi^{0})}(q,k) & = & \int d^{4}x e^{iq.x}\int d^4y
e^{-ik.y}
\langle\Omega\,\vert
T\{V_{\mu}^{\mbox{\tiny em}}(x)V_{\rho}^{\mbox{\tiny
em}}(y)J^{\Phi^{0}}(0)\}\,\vert
\,\Omega\rangle\,, \nn \\
 & = & \epsilon_{\mu\rho\alpha\beta}
\,q^{\alpha}(-k^{\beta})\cH^{\Phi^{0}}\left(q^2,k^2,(q-k)^2 \right)\,,
\eea}

\noi
and $J^{\Phi^{0}}$ is the current defined in Eq.~\rf{unH2}, where the
quark masses enter, linearly, as explicit couplings.
It is the third term in Eq.~\rf{fullWIA} that we shall now be
concerned with. This term, when added
together with the contribution from the unphysical Higgs coupling in
Eqs.~\rf{unH1} and
\rf{unH2} produces an extra contribution to the muon anomaly
$$
F_{2}(0)\big\vert_{\mbox{\rm\tiny long
}}=\frac{g^2}{16\cos^2\theta_{W}}
\frac{e^2}{M_{Z}^2}
\lim_{k^2\ra
0}\int\frac{d^4q}{(2\pi)^4}\frac{1}{q^2}
\times
$$
$$
\frac{m_{\mu}}{2k^2}\tr\left\{(\pslsh+m_{\mu})
\left[\gamma^{\rho}\ksls-\left(k^{\rho}+
\frac{p^{\rho}}{m_{\mu}}\ksls\right)\right]
\left[\gamma^{\mu}\frac{(\pslsh\ -\qsls+m_{\mu})}
{q^2-2q\dd p}
\gamma_{5}+\gamma_{5}
\frac{(\pslsh\ +\qsls+m_{\mu})}{q^2+2q\dd p}\gamma^{\mu}
\right]
\right\}\times
$$
\be\lbl{intlong}
\left[\frac{2M_{Z}}{q^2}\
2\!\!\!\sum_{q=u,d,s}a_{q}\frac{m_{q}}{M_{Z}}
\int d^{4}x e^{iq.x}\int d^4y e^{-ik.y}
\langle\Omega\,\vert
T\{V_{\mu}^{\mbox{\tiny em}}(x)V_{\rho}^{\mbox{\tiny em}}(y)\
\bar{q}(0)i\gamma_{5}q(0)\}\,\vert
\,\Omega\rangle\right]\,,
\ee
which we are next going to evaluate.

At the level of accuracy that we are interested in, it is sufficient
to consider the contribution from the strange quark in the sum of
light quark flavours in Eq.~\rf{intlong}, and let $m_{u}=m_{d}\ra 0\,.$
Then, the relevant hadronic three--point function, in the notation of
ref.~\cite{KN01} is

{\setl
\bea
\left(\Pi_{VVP}\right)_{\mu\rho}(q,-k) & = &\int d^{4}x e^{iq.x}\int
d^4y e^{-ik.y}
\langle\Omega\,\vert
T\{V_{\mu}^{\mbox{\tiny em}}(x)V_{\rho}^{\mbox{\tiny em}}(y)\
\bar{s}(0)i\gamma_{5}s(0)\}\,\vert
\,\Omega\rangle \nn \\
 & = &
\epsilon_{\mu\rho\alpha\beta}q^{\alpha}(-k^{\beta})\left(\frac{4}{9}
\right)
\cH_{V}\left(q^2,k^2,(q-k)^2\right)\,.
\eea}

\noi
In the MHA to large--$N_c$ QCD, it is found~\cite{KPPdeR99,KN01} that:
\be\lbl{MHAVVP}
\cH_{V}^{\mbox{\rm\tiny
MHA}}\left(q^2,k^2,(q-k)^2\right)=-\frac{\stern_{0}}{2}
\frac{q^2+k^2+(q-k)^2-\frac{N_c}{4\pi^2}\frac{M_V^4}{F_{0}^2}}
{(q^2-M_V^2)(k^2-M_V^2)[(q-k)^2-\tilde{M}^2]}\,,
\ee
where, in the denominator, we have kept a pseudoscalar Goldstone mass
$\tilde{M}^2$ in order to handle the infrared logarithmic
divergence which, otherwise, would appear after integration over the
$q$-momentum in \rf{intlong}. Here, we are concerned with the limit of
Eq.~\rf{MHAVVP} when
$k$ is soft. Then
\be
\lim_{k\ra 0}\cH_{V}^{\mbox{\rm\tiny
MHA}}\left(q^2,k^2,(q-k)^2\right)=
\frac{\stern_{0}}{M_V^2 - \tilde{M}^2}\left\{\frac{1-\frac{N_c}
{8\pi^2}\frac{M_V^2}{F_0^2}}{q^2-M_V^2} +
\frac{\frac{N_c}{8\pi^2}\frac{M_V^2}{F_0^2}-
\frac{\tilde{M}^2}{M_V^2}}{q^2-\tilde{M}^2}\right\}\,.
\ee
The integral over the momentum $q$ can now be made, very much the same
way as the calculation of the anomalous contribution in
Eq.~\rf{anomal}, with the result
$$
F_2(0)\vert_{\mathrm{long}}=- \frac{G_F}
{\sqrt 2}\frac{m_\mu^2}{8\pi ^2}\frac{\alpha}{\pi}\
\frac{4}{9} \frac{16 \pi^2 m_s \stern_{0}}{M_V^4}
\frac{1}{1-\frac{\tilde{M}^2}{M_V^2}}
\int_0^1 dx x^2 \int_0^1 dy y \times
$$
$$
\left\{\left(1-\frac{N_c}{8\pi^2}\frac{M_V^2}{F_0^2}\right)
\frac{1}{x(1-y)
+\frac{m_\mu^2}{M_V^2}(1-x)^2}\left(1-\frac{m_\mu^2}{M_V^2}
\frac{(1-x)^2}{x(1-y) +\frac{m_\mu^2}{M_V^2}(1-x)^2}\right)\right.
$$
\be
\left.+
\left(\frac{N_c}{8\pi^2}\frac{M_V^2}{F_0^2}-\frac{\tilde{M}^2}{M_V^2}
\right)\frac{M_V^2}{\tilde{M}^2}
\frac{1}{x(1-y)
+\frac{m_\mu^2}{\tilde{M}^2}(1-x)^2}\left(1-\frac{m_\mu^2}{\tilde{M}^2}
\frac{(1-x)^2}{x(1-y)
+\frac{m_\mu^2}{\tilde{M}^2}(1-x)^2}\right)\right\}\,.
\ee

In the approximation where $m_{\mu}^2\ll
M_{V}^2\,, \ m_{\mu}^2\ll
\tilde{M}^2\,,$ and keeping only logarithmic terms and constant terms, we
get
$$
F_2(0)\vert_{\mathrm{long}}=\frac{G_F}{\sqrt 2}
\frac{m_\mu^2}{8\pi^2}\frac{\alpha}{\pi}\
\frac{4}{9}\frac{8\pi^2F_0^2}{M_V^2}\times
$$
\be
\left\{\frac{N_c}{8\pi^2}\frac{M_V^2}{F_0^2}\ \frac{M_K^2}{\tilde{M}^2}
\left(\log\frac{\tilde{M}^2}{m_{\mu}^2}+\frac{1}{2} \right)+
\left[1-\frac{N_c}{8\pi^2}\frac{M_V^2}{F_0^2}\right]\
\frac{M_K^2}{M_V^2}\frac{M_V^2}{M_{V}^2-\tilde{M}^2}\log\frac{M_{V}^2}
{\tilde{M}^2}\right\}\,,
\ee
where we have used the current algebra relation $m_s
\stern_0
\simeq -F_{0}^2 M_{K}^2\,.$ For the numerical evaluation, we take
$F_0=0.087~\GeV$,
$M_{K}=0.498~\GeV$, $M_{V}=0.770~\GeV$ and let $\tilde{M}$ vary in the
range
$\frac{2}{3} M_K^2\le
\tilde{M}^2\le \frac{4}{3} M_{K}^2$, which includes the values of the
$\eta$ and
$\eta'$ masses induced by $m_s\not= 0$. This results in the value
\be\lbl{longnonan}
F_{2}(0)\big\vert_{\mbox{\rm\tiny long
}}   =   \frac{\GF}{\sqrt{2}}\frac{m_{\mu}^2}{8\pi^2}
\frac{\alpha}{\pi}\times(4.57\pm 1.17\pm 1.37)\,,
\ee
where the first error is the one from the variation of $\tilde{M}$ and the
second one is a
$30\%$ systematic error from the MHA approach.
This result represents a {\it positive} contribution, smaller than the
contribution from the anomaly, but much larger than the one from the
transverse part of the hadronic
$\langle V\!AV\rangle$ Green's function in the chiral limit.

Finally, we still have to consider possible chiral corrections in the
evaluation of the transverse term $\tilde{W}_{\mu\nu\rho}^{\mbox{\rm\tiny
trans }}(q,k)$; i.e., terms linear in quark
masses which may contribute to the invariant function $W(Q^2)$
in Eq.~\rf{intW}. The chiral corrections which are potentially important
are those which modify the asymptotic behaviour of the function
$W(Q^2)$ in Eq.~\rf{opefall} to a less sharp behaviour. The calculation of
the leading behaviour, $1/Q^4$  in this case, is technically rather
involved and we have relegated it to the Appendix C, where it is shown
that
\be\lbl{massOPE}
\lim_{Q^2\ra\infty}W(Q^2)=-\frac{4}{9}\frac{1}{M_{\rho}^2}
\frac{(4m_{u}-m_{d}-m_{s})
\stern_{0}}{Q^4}+
\frac{16}{9}\pi^2\frac{\als}{\pi}\frac{1}{M_{\rho}^2}\frac{\stern_{0}^2}{Q^6}\,,
\ee
where chiral corrections to the $\cO(1/Q^6)$ term and higher order
terms have been neglected.

With $ W(Q^2)=\frac{1}{M_{\rho}^2}w(z)\quad\annd\quad
z\equiv\frac{Q^2}{M_{\rho}^2}\,, $ we now have to construct the
corresponding MHA to large--$N_c$ QCD for the meromorphic
function $w(z)$. From the leading OPE behaviour above, it follows
that the minimum number of poles required is now $p=2$. The
resulting $w(z)$ function is then \be\lbl{mha4}
w(z)\big\vert_{\mbox{\rm\tiny
MHA}}=\frac{\cR_{m}}{(z+1)(z+\frac{1}{g_{A}})}\,,\quad\with\quad
\cR_{m}=-\frac{4}{9}\frac{(4m_{u}-m_{d}-m_{s})
\stern_{0}}{M_{\rho}^4}\,, \ee which leads to an anomalous
magnetic moment contribution
$$
F_{2}(0)\big\vert_{\mbox{\rm\tiny
trans}}^{\mbox{\rm\tiny MHA}}=-\frac{\GF m_{\mu}^2}{\sqrt{2}}
\frac{\alpha}{2\pi}\frac{g_{A}}{1-g_{A}}
\log\left(\frac{1}{g_{A}}\right)\cR_{m}\,.
$$
Neglecting $m_{u}$ and $m_{d}$ with respect to $m_s$, and using
again the current algebra relation
$-m_{s}\stern_{0}\simeq F_{0}^2 M_{K}^2\,,$ we obtain (with $g_{A}=1/2\,,\
F_{0}=0.087~\GeV\,,\  M_{K}=0.498~\GeV$)
\be\lbl{mham}
F_{2}(0)\big\vert_{\mbox{\rm\tiny
trans}}^{\mbox{\rm\tiny
MHA}}\simeq\frac{\GF}{\sqrt{2}}\frac{m_{\mu}^2}{8\pi^2}
\frac{\alpha}{\pi}\
\log2\ \frac{16\pi^2
F_{0}^2}{9M_{\rho}^2}\frac{M_{K}^2}{M_{\rho}^2}=
\frac{\GF}{\sqrt{2}}\frac{m_{\mu}^2}{8\pi^2}
\frac{\alpha}{\pi}\times 0.065\,.
\ee

In this case, however, we know that the MHA in Eq.~\rf{mha4} fails to
reproduce the $\cO(1/Q^6)$ term in the OPE, which we know is there, even
in the chiral limit. In order to take into account this information, we
can go now one step further beyond the MHA and incorporate three poles,
like in Eq.~\rf{mha3poles} with the residues constrained now by
short--distance behaviour as follows:
{\setl
\bea
\alpha_{1}+\beta_{1}+\alpha_{2} & = &0\,, \nn
\\
\alpha_{1}+\frac{1}{g_{A}}\beta_{1}+r'\alpha_{2} & = & -\cR_{m}\,, \nn \\
\alpha_{1}+\frac{1}{g_{A}^2}\beta_{1}+r'^{2}\alpha_{2} & = & \cR \nn
\,.
\eea}

\noi The resulting $w(z)$ function in this improved case is shown
in Fig.~6, in $10^{-3}$ units (the continuous curve). For the
sake of comparison, we also show in the same figure and in the
same units the curve corresponding to the MHA in Eq.~\rf{mha4}
(the dashed curve). The integral over $z$ in the improved case
can still be done analytically and, neglecting terms of
$\cO\left(\frac{m_{\mu}^2}{M_{Z}^2}\log\frac{M_{Z}^2}{m_{\mu}^2}
\right)$, we get

{\setl
\bea\lbl{3poles}
F_{2}(0)\big\vert_{\mbox{\rm\tiny
trans}}^{\mbox{\rm\tiny HA 3--poles}} & = & -\frac{\GF
m_{\mu}^2}{\sqrt{2}}
\frac{\alpha}{2\pi}
\left\{\frac{\left[\cR_{m}+(\cR+\cR_{m})g_{A}\right]\log{r'}}{(1-g_{A}r')
(r'-1)}+
\frac{\left[\cR+(1+r')\cR_{m} \right]g_{A}^2
\log{g_{A}}}{(1-g_{A})(1-g_{A}r')} \right\} \nn \\
 & = & \frac{\GF}{\sqrt{2}}\ \frac{m_{\mu}^2}{8\pi^2}\ \frac{\alpha}{\pi}
\times (0.04\pm 0.02)\,.
\eea}

\noi
As expected, the effect of including the $1/Q^6$ terms in the asymptotic
behaviour of the $W(Q^2)$ function reduces the integral, but the sign is
still the same as in the MHA approximation result in Eq.~\rf{mham}. The
final error we quote is a very generous estimate of the systematic errors.

\vskip 2pc \centerline{\epsfbox{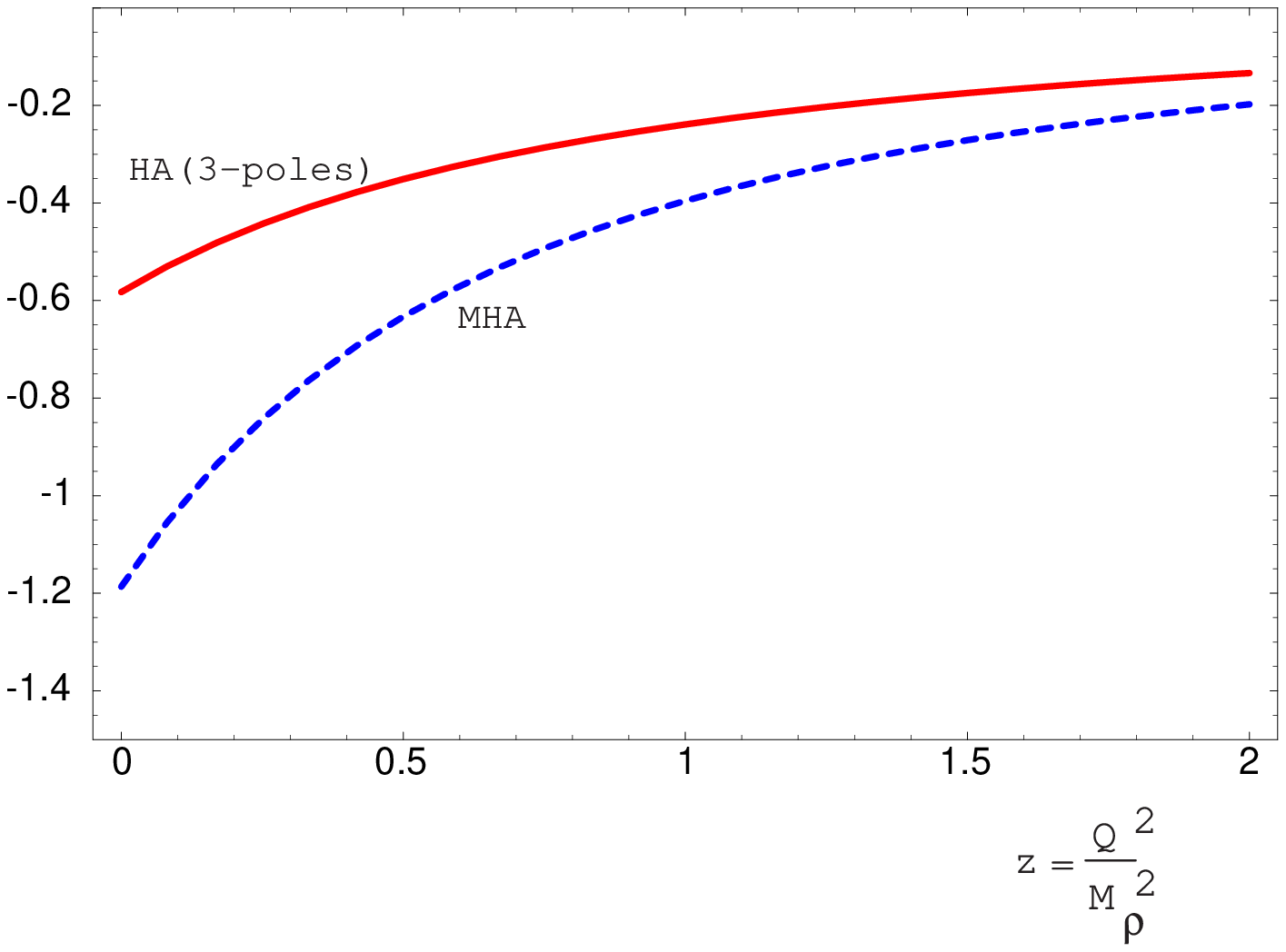}} \vspace*{1cm} {\bf
Fig.~6} {\it Shape of the functions $w(z)$ (in $10^{-3}$ units) in
the hadronic approximation to large--$N_c$ with 3--poles (the
continuous curve) and in the minimal hadronic approximation with
2--poles in Eq.~\rf{mha4} (the dashed  curve).} \vskip 2pc

\vspace{0.7cm}

\section{\normalsize Results and Conclusions}
\setcounter{equation}{0}

\noi
The normalization reference of the electroweak contributions to the
muon anomaly is the one--loop
result~\cite{BGL72,ACM72,JW72,BY72,FLS72}
$$
a_{\mu}^{\mbox{\rm\tiny
W(1)}}=\frac{\GF}{\sqrt{2}}\,\frac{m_{\mu}^2}{8\pi^2}\,
\left[\frac{5}{3}+\frac{1}{3}
\left(1-4\sin^2\theta_{W}\right)^2+
\cO\left(\frac{m_{\mu}^2}{M_{Z}^2}\log\frac{M_{Z}^2}{m_{\mu}^2}
\right)+\cO\left(\frac{m_{\mu}^2}{M_{H}^2}\log\frac{M_{H}^2}{m_{\mu}^2}
\right)
\right]\,,
$$
where the weak mixing angle is defined by
$\sin^2\theta_{W}=1-M_{W}^2/M_{Z}^2\,.$ Numerically, with $\GF=1.166
39(1)\times 10^{-5}\,\GeV^{-2}$ and  $\sin^2\theta_{W}=0.224$,
$$
a_{\mu}^{\mbox{\rm\tiny
W(1)}}=19.48\times 10^{-10}\,,
$$
while we recall that the present {\it world average} experimental error
in the determination of the muon anomaly is
$$
\Delta a_{\mu}\big\vert_{\mbox{\rm\tiny Exp.}}=\pm 15.1\times
10^{-10}\,,
$$
and, hoping for a continuation of the BNL experiment, it is expected to
be further reduced by still a factor of four. A theoretical effort on
the evaluation of the two--loop electroweak corrections is therefore
justified.

It is convenient to separate the two--loop electroweak contributions into
two sets of Feynman graphs: those which contain closed fermion loops,
which we denote by $a_{\mu}^{EW(2)}({\mbox{\rm\footnotesize ferm}})$,
and the others which we denote by
$a_{\mu}^{EW(2)}({\mbox{\rm\footnotesize bos}})$. In this notation,
the electroweak contribution to the muon anomalous magnetic moment is
\be
a_{\mu}^{EW}=a_{\mu}^{W(1)}+a_{\mu}^{EW(2)}({\mbox{\rm\footnotesize
bos}})+a_{\mu}^{EW(2)}({\mbox{\rm\footnotesize ferm}})\,.
\ee
We shall review the calculation of these two--loop contributions
separately.

\subsection{\it Bosonic Contributions}

The leading logarithmic terms of the two--loop electroweak bosonic
corrections have been extracted using asymptotic expansion
techniques~(see e.g. ref.\cite{Smi94}.) In fact, these
contributions have now been evaluated analytically, in a
systematic expansion in powers of $\sin^2\theta_{W}$, up to
$\cO[(\sin^2\theta_{W})^3]\,,$ where
$\log\frac{M_{W}^2}{m_{\mu}^2}$ terms,
$\log\frac{M_{H}^2}{M_{W}^2}$ terms, $\frac{M_{W}^2}{M_{H}^2}
\log\frac{M_{H}^2}{M_{W}^2}$ terms, $\frac{M_{W}^2}{M_{H}^2}$
terms and constant terms are kept~\cite{CKM96}. Using
$\sin^2\theta_{W}=0.224$ and $M_{H}=250\,\GeV\,,$ the authors of
ref.~\cite{CKM96} find \be\lbl{bos}
a_{\mu}^{EW(2)}({\mbox{\rm\footnotesize
bos}})=\frac{\GF}{\sqrt{2}}\,\frac{m_{\mu}^2}{8\pi^2}\,
\frac{\alpha}{\pi}\times
\left[-5.96\log\frac{M_{W}^2}{m_{\mu}^2}+0.19\right]=
\frac{\GF}{\sqrt{2}}\,\frac{m_{\mu}^2}{8\pi^2}\,
\frac{\alpha}{\pi}\times (-78.9)\,. \ee

\subsection{\it Fermionic Contributions}

The discussion of the two--loop electroweak fermionic corrections is
more delicate. As already mentioned in the introduction, because of the
cancellation between lepton loops and quark loops in the electroweak
$U(1)$ anomaly, one cannot separate hadronic effects from leptonic
effects any longer. In fact, as discussed in refs.~\cite{PPdeR95,CKM95},
it is this cancellation which eliminates some of the large logarithms
which,  were incorrectly kept in ref.~\cite{KKSS92}. It is therefore
appropriate to separate the two--loop electroweak fermionic corrections
into two classes: One is the class arising from Feynman diagrams like
in Fig.~1, with both leptons and quarks in the shaded blob, including
the graphs  where the $Z$ lines are replaced by $\Phi^{0}$ lines, if
the calculation is done in the $\xi_{Z}$--gauge.  We denote this class
by
$a_{\mu}^{EW(2)}(l,q)\,.
$ The other class is defined by the rest of the diagrams, where quark
loops and lepton loops can be treated separately, which we call it
$a_{\mu}^{EW(2)}({\mbox{\rm\footnotesize ferm-rest}})$
i.e.,
$$
a_{\mu}^{EW(2)}({\mbox{\rm \footnotesize
fer}})=a_{\mu}^{EW(2)}(l,q)+
a_{\mu}^{EW(2)}({\mbox{\rm\footnotesize ferm-rest}})\,.
$$
The contribution from $a_{\mu}^{EW(2)}({\mbox{\rm\footnotesize
ferm-rest}})$ brings in $m_{t}^2/M_{W}^2$ factors. It has been
estimated, to a very good approximation, in ref.~\cite{CKM95} with the
result
$$
a_{\mu}^{EW(2)}({\mbox{\rm\footnotesize
ferm-rest}})=\frac{\GF}{\sqrt{2}}\,\frac{m_{\mu}^2}{8\pi^2}\,
\frac{\alpha}{\pi}\times\left[
\frac{1}{2\sin^2\theta_{W}}\left(-\frac{5}{8}
\frac{m_{t}^2}{M_{W}^2}-\log\frac{m_{t}^2}{M_{W}^2}-\frac{7}{3}\right)
+\Delta_{\mbox{\rm\tiny Higgs}}\right]\,,
$$
where $\Delta_{\mbox{\rm\tiny Higgs}}$ denotes the contribution from
diagrams with Higgs lines, which the authors of ref.~\cite{CKM95}
estimate to be
$$
\Delta_{\mbox{\rm\tiny Higgs}}=-5.5\pm 3.7\,,
$$
and therefore,
\be\lbl{ferrest}
a_{\mu}^{EW(2)}({\mbox{\rm\footnotesize
ferm-rest}})=\frac{\GF}{\sqrt{2}}\,\frac{m_{\mu}^2}{8\pi^2}\,
\frac{\alpha}{\pi}\times\left(-21\,\pm\,4 \right)\,.
\ee

Let us finally discuss the contributions to
$a_{\mu}^{EW(2)}(l,q)$ which are the ones
relevant to the topic of this paper. Here, it is convenient to treat
the contributions from the three generations separately. The
contribution from the third generation can be calculated in a
straightforward way, with the result~\cite{PPdeR95,CKM95}

{\setl
\bea\lbl{3rdg}
a_{\mu}^{EW(2)}(\tau,t,b) & = & \frac{\GF}{\sqrt{2}}\,\frac{m_{\mu}^2}
{8\pi^2}\,
\frac{\alpha}{\pi}  \times
 \left[-3\log\frac{M_{Z}^2}{m_{\tau}^2}-\log\frac{M_{Z}^2}{m_{b}^2}-
\frac{8}{3}\log\frac{m_{t}^2}{M_{Z}^2}+\frac{8}{3} +
\cO\left(\frac{M_{Z}^2}{m_{t}^2}\log\frac{m_{t}^2}{M_{Z}^2}
\right)
\right] \nn \\
 &= & \frac{\GF}{\sqrt{2}}\,\frac{m_{\mu}^2}
{8\pi^2}\,
\frac{\alpha}{\pi}\times (-30.6)\,.
\eea}

\noi
In fact the terms of $\cO\left(\frac{M_{Z}^2}{m_{t}^2}
\log\frac{m_{t}^2}{M_{Z}^2}
\right)$ and  $\cO\left(\frac{M_{Z}^2}{m_{t}^2}\right)$ have also been
calculated in ref.~\cite{CKM95}. There are in principle QCD
perturbative corrections to this estimate, which have not been
calculated, but the result in Eq.~\rf{3rdg} is good enough for the
accuracy required at present.

As emphasized in ref.~\cite{PPdeR95}, an appropriate QCD calculation
when the quark in the loop of Fig.~1 is a {\it light quark} should
take into account the dominant effects of spontaneous chiral symmetry
breaking. Since this involves the $u$ and $d$ quarks, as well as the
second generation $s$ quark, we lump together the contributions  from
the first and second generation with the result,

{\setl
\bea\lbl{12g}
a_{\mu}^{EW(2)}(e,\mu,u,d,s,c) &=
&\frac{\GF}{\sqrt{2}}\,\frac{m_{\mu}^2} {8\pi^2}\,
\frac{\alpha}{\pi}  \times  \left\{-3\log
\frac{M_{Z}^2}{m_{\mu}^2} -\frac{5}{2}
\right. \nn \\
& &  \ \ \ \ \ \ \ \ \ \ \ \ \ \ \ \ \ \ \  \
-3\log\frac{M_{Z}^2} {m_{\mu}^2}+4\log\frac{M_{Z}^2}{m_{c}^2}
-\frac{11}{6}+\frac{8}{9}\pi^2-8
\nn
\\ &+ &\! \left.
\left[\frac{4}{3}\log\frac{M_{Z}^2}{m_{\mu}^2}\!+\!\frac{2}{3}\!+\!\cO\left(
\frac{m_{\mu}^2}{M_{Z}^2}\log\frac{M_{Z}^2}{m_{\mu}^2}
\right)\!\right]\! +\!4.57\pm 1.80 +0.04\pm 0.02\ \right\}\\
& = & \lbl{12gs}\frac{\GF}{\sqrt{2}}\,\frac{m_{\mu}^2} {8\pi^2}\,
\frac{\alpha}{\pi} \times (-28.5\pm 1.8)\,,
\eea}

\noi
where the first line in Eq.~\rf{12g} shows
the result from the $e$--loop and the second line the result from the
$\mu$--loop and the $c$--quark, which is treated as a heavy quark ($m_c
=1.5~\GeV$). The first term in brackets in the third line is the one
induced by the anomalous term in the hadronic
$<V\!AV>$ Green's function.
The second term is the one induced by the leading
effects of explicit chiral symmetry breaking in the
non--anomalous longitudinal component of the $<V\!AV>$ Green's
function discussed in section~6 (the result in Eq.~\rf{longnonan}). The
third term is the one induced by the transverse component of the
$<V\!AV>$ Green's function, evaluated in the presence of the light quark
masses and in the HA with 3 poles to large--$N_c$ QCD (the result in
Eq.~\rf{3poles}).

We want to stress that our result in Eq.~\rf{12g} for the
contribution from the first and second generations of quarks and leptons
is conceptually very different to the corresponding one proposed in
ref.~\cite{CKM95},

{\setl
\bea\lbl{12gCKM}
a_{\mu}^{EW(2)}(e,\mu,u,d,s,c) & = &
\frac{\GF}{\sqrt{2}}\,\frac{m_{\mu}^2} {8\pi^2}\,
\frac{\alpha}{\pi}
\left[-3\log\frac{M_{Z}^2}
{m_{\mu}^2}+4\log\frac{M_{Z}^2}{m_{u}^2}-\log\frac{M_{Z}^2}{m_{d}^2}
-\frac{5}{2}-6
\right. \nn \\
& & \ \ \ \ \ \ \ \ — \  \,\left.
-3\log\frac{M_{Z}^2}
{m_{\mu}^2}+4\log\frac{M_{Z}^2}{m_{c}^2}-\log\frac{M_{Z}^2}{m_{s}^2}
-\frac{11}{6}+\frac{8}{9}\pi^2-6\right]\,, \nn\\
& = & \frac{\GF}{\sqrt{2}}\,\frac{m_{\mu}^2} {8\pi^2}\,
\frac{\alpha}{\pi} \times (-31.9)\,.
\eea}

\noi where the light quarks are, {\it arbitrarily},  treated the
same way as heavy quarks, with $m_{u}=m_{d}=0.3\,\GeV\,,$ and
$m_{s}=0.5\,\GeV\,.$ Although, numerically, the two results turn
out to be not too different, the result in Eq.~\rf{12gCKM}
follows from a hadronic model which is in contradiction with
basic properties of QCD. The constituent quark model used to
derive this result violates the current algebra Ward
identity~\rf{cawi} derived in Appendix A.  Furthermore, as
discussed in the text, the model does not reproduce the QCD
short--distance behaviour for the underlying $<V\!AV>$ Green's
function. These facts are at the origin of the
spurious cancellation of the $\log M_{Z}$ terms in
Eq.~\rf{12gCKM}. A more detailed discussion of this
issue can be found in Appendix D.

Putting together the numerical results in Eqs.~\rf{bos}, \rf{ferrest},
\rf{3rdg} with the new result in Eq.~\rf{12gs}, we finally obtain the
value
$$
a_{\mu}^{EW}=\frac{\GF}{\sqrt{2}}\,\frac{m_{\mu}^2}
{8\pi^2}\left[\frac{5}{3}+\frac{1}{3}
\left(1-4\sin^2\theta_{W}\right)^2-\left( \frac{\alpha}{\pi}\right)
(159\,\pm\,4)\right]=(15.2\, \pm\, 0.1)\times 10^{-10}\,,
$$
which shows  that the two--loop correction represents indeed a reduction
of the one--loop result by an amount of $22\%\,.$ The final error here
does not include higher order electroweak estimates~\cite{DG98}.

\vspace*{2cm}

\begin{center}
{\Large\bf Acknowledgements}
\end{center}

\noi This work has been partly supported by TMR, EC--Contracts No.
ERBFMRX-CT980169 (EURO\-DA$\Phi$NE), HPRN-CT-2002-00311
(EURIDICE) and the France-Spain action int\'{e}gr\'{e}e
``Picasso''. The work of S.P. is also supported by
CICYT-AEN99-0766, CICYT-FEDER-FPA2002-00748 and 2001 SGR00188.


\newpage
\appendix
\begin{center}
{\Large\bf APPENDIX}
\end{center}

\section{\large Ward Identities and Gauge Invariance}
\setcounter{equation}{0}
\def\theequation{\Alph{section}.\arabic{equation}}

The hadronic Green's function in Eq.~\rf{VAV} is a combination of QCD
three--point functions involving the vector and axial vector currents
$$
V^a_{\mu}(x)\,=\,
\left({\overline\psi}\,\frac{\lambda^a}{2}\,
\gamma_{\mu}\psi\right)(x)\,,\qquad\annd\qquad
\ A^a_{\mu}(x)\,=\,
\left({\overline\psi}\,\frac{\lambda^a}{2}\,
\gamma_{\mu}\gamma_5\psi\right)(x)\,,
$$
of the  algebra of currents, where
$\lambda^a$ denotes the flavour Gell-Mann matrices defined after
Eq.~\rf{VTfunc}. More precisely, with
$$
{\cal T}^a_{\mu\nu;\alpha}(q_1,q_2)\ =\ i\int d^4x_1 d^4x_2
\,e^{i(q_1\cdot x_1 + q_2\cdot x_2)}
\,<\,\Omega\,\vert\,\mbox{T}\{V_\mu^{\mbox{\tiny
em}}(x_1)V_\nu^{\mbox{\tiny
em}}(x_2)A^a_\alpha(0)\}
\,\vert\,\Omega\,>\,,
$$
where
$$
V_{\mu}^{\mbox{\tiny
em}}(x) \,=\, V^{3}_{\mu}(x)\,+\,\frac{1}{\sqrt{3}}\,V^{8}_{\mu}(x)\,,
$$
the relation is the following
$$
\frac{i}{2}\,W_{\mu\alpha\nu}(q,k)\,=\,
{\cal T}^{3}_{\mu\nu;\alpha}(q,-k)\,+
\,\frac{1}{\sqrt{3}}\,{\cal T}^{8}_{\mu\nu;\alpha}(q,-k)\,-
\,\frac{1}{\sqrt{6}}\,{\cal T}^{0}_{\mu\nu;\alpha}(q,-k)\,.
$$

\subsection{\em Current Algebra Ward Identities}

For $a=0,3,8$, one has the following Ward
identities~\footnote{Recall that we are using
the following conventions: $\epsilon_{0123}\,=\,+1$ and
$\gamma_5 = i\gamma^0\gamma^1\gamma^2\gamma^3$, so that
$\mbox{tr}(\gamma_5\gamma_0\gamma_1\gamma_2\gamma_3) = 4i$.}:

{\setl\bea\lbl{cawi}
(q_1 +q_2)^\alpha\,{\cal T}^a_{\mu\nu;\alpha}(q_1,q_2) &=&
-\ \frac{N_c}{12\pi^2}\,{\cal C}^a\,
\epsilon_{\mu\nu\rho\sigma}q_1^\rho q_2^\sigma\,
\nonumber\\
&+ & \int d^4x_1 d^4x_2
\,e^{i(q_1\cdot x_1 + q_2\cdot x_2)}
\,<\Omega\,\vert\,\mbox{T}\{V_\mu^{\mbox{\tiny
em}}(x_1)V_\nu^{\mbox{\tiny
em}}(x_2)D^a_5(0)\}
\,\vert\,\Omega>
\nonumber\\
+ \,\frac{1}{8\pi^2}\,\mbox{tr}\left(\frac{\lambda^a}{2}\right)
  &\times&\int d^4x_1 d^4x_2\,e^{i(q_1\cdot x_1 + q_2\cdot x_2)}
<\Omega\vert\mbox{T}\{V_\mu^{\mbox{\tiny
em}}(x_1)V_\nu^{\mbox{\tiny
em}}(x_2)(G{\widetilde G})(0)\}
\vert\Omega>\,,
\eea}

\noi
with ${\cal C}^3=1$, ${\cal C}^8=1/\sqrt{3}$, ${\cal C}^0=2\sqrt{2/3}$.
Furthermore,
$$
D^a_5\,\equiv\,d^{abc}{\cal M}^bP^c\,,\quad
\ {\mbox{with}}\quad
\ P^a\,=\,{\overline\psi}i\gamma_5\,\frac{\lambda^a}{2}\,\psi\,,
$$
and
\bea
\ {\cal M}^3 = m_u-m_d\,,\quad
\ {\cal M}^8 = \frac{1}{\sqrt{3}}(m_u+m_d-2m_s)\,,\quad
\ {\cal M}^0 = \sqrt{\frac{2}{3}}(m_u+m_d+m_s)\,.
\nonumber
\eea
Also
$$
(G{\widetilde
G})=\alpha_s\,\epsilon_{\mu\nu\rho\sigma}
\sum_{A=1}^{8}G^{(A)\mu\nu}G^{(A)\rho\sigma}\,,
$$
with $G^{(A)\mu\nu}$ the gluon field strength tensor.
Owing to the fact that

{\setl\bea
\{q_1^{\mu};q_2^{\nu}\}\,\int d^4x_1 d^4x_2
\,e^{i(q_1\cdot x_1 + q_2\cdot
x_2)}\,<\Omega\,\vert\,\mbox{T}\{V_\mu^{\mbox{\tiny
em}}(x_1)V_\nu^{\mbox{\tiny
em}}(x_2)(G{\widetilde
G})(0)\}
\,\vert\,\Omega>\ & = & \ 0\,,
\nonumber\\
\{q_1^{\mu};q_2^{\nu}\}\,\int d^4x_1 d^4x_2
\,e^{i(q_1\cdot x_1 + q_2\cdot x_2)}\,
<\Omega\,\vert\,\mbox{T}\{V_\mu^{\mbox{\tiny
em}}(x_1)V_\nu^{\mbox{\tiny
em}}(x_2)D^a_5(0)\}
\,\vert\,\Omega>\ &  = & \ 0\,,
\nonumber
\eea}

\noi
the above Ward identity can be solved as follows:

{\setl
\bea\lbl{fullWIA}
\!\!\!\!\!\!\!\!\!{\cal T}^a_{\mu\nu;\alpha}(q_1,q_2) &=&
\frac{(q_1+q_2)_{\alpha}} {(q_1+q_2)^2}\,
\Big\{
-\, \frac{N_c}{12\pi^2}\,{\cal C}^a\,
\epsilon_{\mu\nu\rho\sigma}q_1^\rho q_2^\sigma
\nonumber\\
&&\!\!\!\!\!
+\ \frac{1}{8\pi^2}\,\mbox{tr}\left(\frac{\lambda^a}{2}\right)
 \int d^4x_1 d^4x_2\,e^{i(q_1\cdot x_1 + q_2\cdot x_2)}
<\Omega\,\vert\,\mbox{T}\{V_\mu^{\mbox{\tiny
em}}(x_1)V_\nu^{\mbox{\tiny
em}}(x_2)(G{\widetilde G})(0)\}
\,\vert\,\Omega>
\nonumber\\
&&\!\!\!\!\!
+\int d^4x_1 d^4x_2
\,e^{i(q_1\cdot x_1 + q_2\cdot x_2)}
<\Omega\,\vert\,\mbox{T}\{V_\mu^{\mbox{\tiny
em}}(x_1)V_\nu^{\mbox{\tiny
em}}(x_2)D^a_5(0)\}
\,\vert\,\Omega>\Big\}
\nonumber\\
&&\!\!\!\!\!
+\ {\overline{\cal T}}^{\,a}_{\mu\nu;\alpha}(q_1,q_2)\, ,
\eea}
where the tensor ${\overline{\cal T}}{\,^a}_{\mu\nu;\alpha}(q_1,q_2)$
 is now completely transverse,
\bea
\{q_1^{\mu};q_2^{\nu};(q_1+q_2)^{\alpha}\}\,
{\overline{\cal T}}^{\,a}_{\mu\nu;\alpha}(q_1,q_2)\ =\ 0\,.
\nonumber
\eea

The first term in Eq.~\rf{fullWIA} generates the anomaly term discussed
in the text. The second term is the one induced by the axial
flavour--singlet
$U(1)$ anomaly. This term generates subleading contributions in the
$1/N_c$--expansion with respect to the other terms; however, because of
its flavour singlet nature, the {\it
numerical size} of its contribution may not necessarily obey the
{\it expected} $1/N_c$--power suppression. However, in our case, the
source of the hadronic axial current is the one induced by the
$Z$--boson coupling in the Standard Electroweak Theory, which has a
vanishing flavour trace for each generation. Therefore, once the total
sum of light and heavy quarks is taken into account, this term does not
contribute to the muon anomaly.

The third term in Eq.~\rf{fullWIA} is the one induced by the
explicit breaking of chiral symmetry by the light quark masses in
the QCD Lagrangian. We keep it, because it will be pertinent in the
forthcoming discussions of electroweak gauge invariance. Notice, for
further reference, that
$$
 D^3_5 \,+\, \frac{1}{\sqrt{3}}\,D^8_5 \,-\,\frac{1}{\sqrt{6}}\,D^0_5
\,=\,
m_u {\bar u}i\gamma_5 u \,-\, m_d {\bar d}i\gamma_5 d \,-
\, m_s {\bar s}i\gamma_5
s\,.
$$

\subsection{\em Electroweak Gauge Invariance}

Due to the conservation of the electromagnetic current, the gauge
dependent terms in the photon propagator in the diagrams of Fig.~1
trigger naive Ward identities on the hadronic $\gamma-\gamma-Z$
vertex, and therefore we can take the Feynman gauge for the photon
propagator without loss of generality, as we
have done in the calculation reported in the text. By contrast, the
gauge dependence of the
$Z$--propagator requires some discussion. In the Electroweak Theory,
it is convenient to rewrite the Ward identity discussed in the previous
subsection in terms of the coupling of the unphysical Higgs field
$\Phi^{0}(x)$. More precisely, in the notation used in the text,
\be\lbl{axialwiA}
(q-k)^{\nu}W_{\mu\nu\rho}(q,k)=-i\frac{N_c}{\pi^2}{\cal
C}_{\mbox{\tiny{had}}}\epsilon_{\mu\rho\alpha\beta}\,
q^{\alpha}k^{\beta}-2i M_{Z}W_{\mu\rho}^{(\Phi^{0})}(q,k)\,.
\ee
where
\bea\lbl{aqeq}
{\cal C}_{\mbox{\tiny{had}}}
\,=\,
 \sum_{q} {\mbox{a}}_q e_q^2 \,,\qquad\with\qquad
\left.\begin{array}{c} {\mbox{a}}_{q}=+1/2\,,\quad
{\mbox{e}}_{q}=+2/3\,,\quad\foor\quad
q=u,c,t \\
{\mbox{a}}_{q}=-1/2\,,\quad{\mbox{e}}_{q}=-1/3\,,\quad\foor\quad q=d,s,b
\end{array}\right\}\,,
\eea
and
$$
W_{\mu\rho}^{(\Phi^{0})}(q,k)=\int d^{4}x e^{iq.x}\int d^4y e^{-ik.y}
\langle\Omega\,\vert
T\{V^{\mbox{\rm\tiny em}}_{\mu}(x)V^{\mbox{\rm\tiny
em}}_{\rho}(y)J^{\Phi^{0}}(0)\}\,\vert
\,\Omega\rangle\,,
$$
with
$$
J^{\Phi^{0}}=2\sum_{q}
{\mbox{a}}_{q}\frac{m_{q}}{M_{Z}}\bar{q}i\gamma_{5}q\,.
$$
This means that in full generality, in the Electroweak Theory,
\be\lbl{EWWI}
W_{\mu\nu\rho}(q,k)=-i\cA\left(q^2,k^2,(q-k)^2 \right)
\frac{(q-k)_{\nu}}{(q-k)^2}\,\epsilon_{\mu\rho\alpha\beta}
\,q^{\alpha}k^{\beta}+\tilde{W}_{\mu\nu\rho}^{\mbox{\tiny trans}}(q,k)\,
\ee
with
\be\lbl{nontr}
\cA\left(q^2,k^2,(q-k)^2 \right)=\frac{N_c}{\pi^2}{\cal
C}_{\mbox{\tiny{had}}}-2M_{Z}\ \cH^{\Phi^{0}}\left(q^2,k^2,(q-k)^2
\right)\,,
\ee
where we have set
$$
W_{\mu\rho}^{(\Phi^{0})}(q,k)=\epsilon_{\mu\rho\alpha\beta}
\,q^{\alpha}(-k^{\beta})\cH^{\Phi^{0}}\left(q^2,k^2,(q-k)^2 \right)\,.
$$

Now, in the calculation corresponding to Fig.~1, the most general form
of the
$Z$--propagator, in a renormalizable linear gauge of the 't~Hooft class,
is

{\setl
\bea
\lefteqn{\cP_{(\xi_{Z})}^{\nu\nu'}(q-k)=
\frac{-i}{(q-k)^2-M_{Z}^2}\left[g^{\nu\nu'}-(1-\xi_{Z})\frac{(q-k)^{\nu}
(q-k)^{\nu'}}{(q-k)^2-\xi_{Z}\,M_{Z}^2}
\right]=} \lbl{prop1}\\
&& \lbl{prop2}
-i\frac{\left[g^{\nu\nu'}(q-k)^2-(q-k)^{\nu}(q-k)^{\nu'}\right]-
\xi_{Z}M_{Z}^2\left(g^{\nu\nu'}-
\frac{(q-k)^{\nu}(q-k)^{\nu'}}{M_{Z}^2}
\right)}{\left[(q-k)^2-M_{Z}^2\right]
\left[(q-k)^2-\xi_{Z}M_{Z}^2\right]}\,.
\eea}

\noi
The unitary gauge, which was used e.g. in ref.~\cite{PPdeR95}, is
recovered in the limit
$\xi_{Z}\ra\infty$, which has to be taken from the start.  In the text
we have done the calculation in the Feynman gauge
$\xi_{Z}=1$. Clearly, there is no contribution to the muon anomaly from
the contraction of the gauge dependent piece in the $Z$--propagator in
\rf{prop1} with the fully transverse $\tilde{W}_{\mu\nu\rho}(q,k)$
hadronic tensor. The contribution from the non--transverse term in
Eq.~\rf{EWWI} however, is gauge dependent. It selects the
second term in the $Z$--propagator in \rf{prop2}, via the contraction
$$
\cP_{(\xi_{Z})}^{\nu\nu'}(q-k)\times\frac{(q-k)_{\nu}}{(q-k)^2}=
\frac{-i\xi_{Z}}{(q-k)^2-\xi_{Z} M_{Z}^2}
\times\frac{(q-k)^{\nu'}}{(q-k)^2}
$$
As shown in Eq.~\rf{nontr} the non--transverse term in Eq.~\rf{EWWI}
has two pieces. Let us first discuss the one coming  from
the anomaly term. For a specific fermion
$f$, the piece coming from the anomaly results in the expression (in
the
$\xi_{Z}$ gauge)
$$
F_2(0)\Big\vert^{(f)}_{\mbox{\tiny anom}} = \frac{G_{\mbox{\tiny
F}}}{\sqrt{2}}\,
\frac{m_{\mu}^2}{8\pi^2}\,\frac{\alpha}{\pi}\,(4e_f^2{\mbox{a}}_f)\times
\left[ \ln\left(\frac{\xi_Z M_Z^2}{m_{\mu}^2}\right)\,+\,
\frac{1}{2}\,+\,{\cal
O}\left(\frac{m_{\mu}^2}{M_Z^2}\,\log\frac{M_Z^2}{m_{\mu}^2}\right)
\right]\,,
$$
while in the unitarity gauge, the corresponding result is
divergent and requires regularization. In dimensional
regularization, with $
\lambda_{\mbox{\tiny{$\overline{\mbox{MS}}$}}}\,\equiv\,
\frac{\mu^{d-4}}{16\pi^2}\,\Big[ \,\frac{1}{d-4}\,-\,
\frac{1}{2}\,\big(\log 4\pi +2+ \Gamma '(1)\big)\Big]\,,$ one
finds
$$
F_2(0)\Big\vert^{(f)}_{\mbox{\tiny anom}} = \frac{G_{\mbox{\tiny
F}}}{\sqrt{2}}\,
\frac{m_{\mu}^2}{8\pi^2}\,\frac{\alpha}{\pi}\,(4e_f^2{\mbox{a}}_f)\times
\left[ \log\left(\frac{\mu^2}{M_{Z}^2}\right)\,
-\,32\pi^2\lambda_{\mbox{\tiny{$\overline{\mbox{MS}}$}}}\,+
\log\left(\frac{M_{Z}^2}{m_{\mu}^2}\right)\,+\, \frac{1}{2}
\right]\,.
$$
Recall that in these expressions, when the fermion $f$ is a quark $q$,
${\mbox{a}}_{q}$ and ${\mbox{e}}_{q}$ have been defined in
Eq.~\rf{aqeq}. When $f$ is a lepton $l$, ${\mbox{a}}_{l}=-1/2$ and
${\mbox{e}}_{l}=-1$.
In particular, for $f=u,d,s\,,$
$$
\sum_{u,d,s} 4e_f^2{\mbox{a}}_f \,=\, 4 N_c\,\frac{1}{2}\,
\bigg[\left(\frac{2}{3}\right)^2\,+\,2\left(\frac{-1}{3}\right)^2(-1)\bigg]
\,=\,\frac{N_c}{3}\,\frac{4}{3}\,,
$$
and in the Feynman gauge where $\xi_{Z}=1$, we recover the result
of Eq.~\rf{anomal}. Notice also that, in the Standard Theory, the
$\xi_{Z}$--gauge dependence, or the
$\log\left(\frac{\mu^2}{M_{Z}^2}\right)\!
-\!32\pi^2\lambda_{\mbox{\tiny{$\overline{\mbox{MS}}$}}}$
dependence in the unitary gauge, cancel when quarks and leptons
are taken together generation by generation.

Next, we discuss the gauge dependent piece coming from the term
proportional to $\cH^{\Phi^{0}}\left(q^2,k^2,(q-k)^2\right)$ in
Eq.~\rf{nontr}, i.e. the term
$$
\cP_{(\xi_{Z})}^{\nu\nu'}(q-k)\times \cH^{\Phi^{0}}
\frac{(q-k)_{\nu'}}{(q-k)^2}=
\cH^{\Phi^{0}}\ \frac{-i\xi_{Z}}{(q-k)^2-\xi_{Z}M_{Z}^2}
\ \frac{(q-k)^{\nu}}{(q-k)^2}\,.
$$
This term, when acting on the leptonic line (see
Eq.~\rf{vertex}) triggers a trivial Ward identity:

{\setl
\bea\lefteqn{
(q-k)_\nu\left\{\bar{u}(p')
\left[\gamma^{\mu}\frac{i}{\pslsh'-\qsls-m_{\mu}}\gamma^{\nu}\gamma_{5}+
\gamma^{\nu}\gamma_{5}\frac{i}{\pslsh+\qsls-m_{\mu}}
\gamma^{\mu}\right]u(p) \right\}=} \nn \\ & &
-2m_{\mu}\bar{u}(p')
\left[\gamma^{\mu}\frac{i}{\pslsh'-\qsls-m_{\mu}}\gamma_{5}+
\gamma_{5}\frac{i}{\pslsh+\qsls-m_{\mu}}
\gamma^{\mu}\right]u(p) \,.
\eea}

\noi
The resulting expression:
$$
\cH^{\Phi^{0}}\
\frac{-i\xi_{Z}}{(q-k)^2-\xi_{Z}M_{Z}^2}\frac{1}{(q-k)^2}\times
2m_{\mu}\bar{u}(p')
\left[\gamma^{\mu}\frac{i}{\pslsh'-\qsls-m_{\mu}}\gamma_{5}+
\gamma_{5}\frac{i}{\pslsh+\qsls-m_{\mu}}
\gamma^{\mu}\right]u(p) \,,
$$
when added together with the gauge dependent term from the
unphysical Higgs contribution to the muon anomaly,
$$
\cH^{\Phi^{0}}\ \frac{i}{M_{Z}^2 [(q-k)^2-\xi_{Z} M_{Z}^2]}
\times 2m_{\mu}\bar{u}(p')
\left[\gamma^{\mu}\frac{i}{\pslsh'-\qsls-m_{\mu}}\gamma_{5}+
\gamma_{5}\frac{i}{\pslsh+\qsls-m_{\mu}}
\gamma^{\mu}\right]u(p)\,,
$$
results in an overall gauge independent contribution, proportional to
\be\lbl{wiresult}
\cH^{\Phi^{0}}\ \frac{i}{M_{Z}^2 (q-k)^2}\times
2m_{\mu}\bar{u}(p')
\left[\gamma^{\mu}\frac{i}{\pslsh'-\qsls-m_{\mu}}\gamma_{5}+
\gamma_{5}\frac{i}{\pslsh+\qsls-m_{\mu}}
\gamma^{\mu}\right]u(p)\,.
\ee

In the unitary gauge, there are no
unphysical Higgs couplings and the piece coming from the term
proportional to $\cH^{\Phi^{0}}$ in Eq.~\rf{nontr}
generates the factor
$$
\cH^{\Phi^{0}}\
\left[-i\frac{g^{\nu\nu'}-\frac{(q-k)^{\nu}(q-k)^{\nu'}}
{M_{Z}^2}}{(q-k)^2-M_{Z}^2}\right]\times
\frac{(q-k)_{\nu'}}{(q-k)^2}=
\cH^{\Phi^{0}}\ \frac{i}{M_{Z}^2(q-k)^2}(q-k)^{\nu}\,,
$$
which, after contraction with the leptonic line, produces an
expression identical to the one in Eq.~\rf{wiresult}.


\vspace*{1cm}

\section{\large Technical Remarks on the OPE in the Chiral Limit}
\setcounter{equation}{0}
\def\theequation{\Alph{section}.\arabic{equation}}

As already mentioned in the text, the relevant OPE is the one of
the product of the two currents
$V_{\mu}^{\elm}(x)A_{\nu}^{\nc}(y)$ in Eq.~\rf{VAV} when $x\ra
y$, because these are the currents where the hard virtual
momentum $q$ can get through. There are two types of
contributions to discuss in the chiral limit.

\subsection{\em Contribution from Soft Gluons}

These are the
contributions generated by the term
\bea\lbl{soft}
\lefteqn{\lim_{q\ra\infty}\int d^4z
e^{iq.z}\ T\left\{V_{\mu}^{\elm}(z)A_{\nu}^{\nc}(0)\right\}\doteq}
\nn
\\  &  &:\bar{q}^i
(0) \gamma_\mu\, Q
\, \ iS^{i j}(q)
\gamma_\nu \gamma_5 Q_{L}^{(3)} \, q^j(0): + :\bar{
q}^j (0)
\gamma_\nu \gamma_5 Q_{L}^{(3)}
\, i \tilde{S}^{j i}(q) \gamma_\mu Q \,
q^i(0) :
\eea
in the perturbative Wick--expansion ($i,j$ denote quark color
indices) when the  short--distance
expansion of the quark propagator~\footnote{See e.g. Eq.~(2.34) in
ref.~\cite{NSVZ84}. Notice that $i\tilde{S}^{ji}(q)=\int d^4z
e^{-iq\cdot z} i\tilde{S}^{ji}(0,z)\,.$} is inserted: {\setl
\bea\lbl{sdeqp}
iS^{i j}(q) & = & \int \! d^4 z\, e^{iq \cdot z}\, iS^{i j}(z,0)
=\delta^{i j}\frac{i}{\not \! q}  \\
&   &
 -
i\frac{q^\alpha}{q^4}g_s {\tilde G}_{\alpha\beta}^{i j}\gamma^\beta
\gamma_5  \nn\\
 &   & +\frac{2i}{3}g_s \frac{1}{q^6}\big[q^2 D^\alpha G_{\alpha
\beta}^{i j}
\gamma^\beta - \not \! q\
D^\alpha G_{\alpha\beta}^{i j}\ q^\beta - q^\gamma D_\gamma\ q^\alpha
G_{\alpha\beta}^{i j}
\gamma^\beta - 3iq^\gamma D_\gamma q^\alpha
{\tilde G}_{\alpha \beta}^{i j}
\gamma^\beta \gamma_5
\big] \lbl{first}\\ &    & +{\cO}\big(\frac{1}{q^5} \big) \nn \,,
\eea}

\noi
where ( $A$ is a color index in the adjoint representation)
$$
G_{\alpha\beta}^{i j} = \frac{1}{2}\sum_{A}\lambda^A_{i
j}G^{(A)}_{\alpha\beta}(0)\,,
\qquad {\tilde G}^{i j}_{\alpha\beta} = \frac{1}{2}
\epsilon_{\alpha \beta \rho\sigma} G^{\rho\sigma}_{i j}\,.
$$
The terms of lowest dimension in the short--distance expansion of
the quark propagator in Eq.~\rf{sdeqp} which can
contribute ({\it a priori}) to a tensor structure like the one in
Eq.~\rf{opeterm} are those with two uncontracted powers of
$q$--momenta in the numerator i.e.,  the last three terms in
Eq.~\rf{first}. However, these terms, as far as their Dirac structure is
concerned, are all proportional to one gamma matrix or to
$\gamma^{\beta}\gamma_{5}$, (the last term.) Therefore, when
inserted in the r.h.s. of  Eq.~\rf{soft}, and taking into account that
$$
Q\ Q_{L}^{(3)}=Q_{L}^{(3)}Q=\frac{1}{3}\mbox{\rm diag}(2,1,1)\,,
$$
plus the identity
\be\lbl{diracid}
\gamma_\mu \gamma_\rho \gamma_\nu =
g_{\mu\rho} \gamma_\nu +
g_{\rho\nu}
\gamma_\mu - g_{\mu\nu}\gamma_\rho -
i \epsilon_{\mu \rho \nu
\delta}\,\gamma^\delta \gamma_5\,,
\ee
one finds that these terms cannot produce an antisymmetric tensor in
the $\mu$ and $\nu$ indices. We then conclude that there is no
contribution from {\it soft--gluons} with the tensor structure of
Eq.~\rf{opeterm} with a power $p\le 3$.

\subsection{\em Contribution from Hard Gluons}

Expanding the $T$--product in Eq.~\rf{Ufunction} in perturbation
theory to $\cO(g_s^2)$ generates the following operator
{\setl
\bea
\cU_{\mu\nu}(q) & = & \frac{i^2}{2!}g_s^2
\lim_{q\ra\infty}\int
d^4z\ e^{iq\cdot z}\int d^4 x\int d^4 y\times \nn \\ & & T\left\{
:\bar{\psi}_{i}(x)\gamma^{\alpha}\frac{\lambda^{A}_{ij}}{2}
G_{\alpha}^{(A)}(x)
\psi_{j}(x):\ :\bar{\psi}_{k}(y)\gamma^{\beta}\frac{\lambda^{B}_{kl}}
{2}G_{\beta}^{(B)}(y)\psi_{l}(y):\right. \nn \\ & &
\left.:\bar{q}(z)\gamma_{\mu}Qq(z):\
:\bar{q}(0)\gamma_{\nu}\gamma_{5}Q_{L}^{(3)}q(0):\right\}\nn\,.
\eea}

\noi
The contraction of the two gluon fields generates four possible
classes of {\it hard} topologies illustrated in Fig.~3 above. Using
the identity
$$
\sum_{A}\frac{\lambda_{i j}^A}{2}
\frac{\lambda_{k l}^A}{2} =
\frac{1}{2}\left(\delta_{i l}\delta_{j k} -\frac{1}{N_c}\delta_{i j}
\delta_{k l}\right)\,,
$$
and the fact that the second term in this identity generates
subleading terms in the $1/N_c$--expansion, the contributions from
the four {\it hard} topologies can be written as follows:

{\setl
\bea\lbl{umlAS}
\cU_{\mu\nu}^{(1)}(q){\Big\vert}_{q\ra\infty}
& = &  -i\frac{i^2}{2!} g_s^2
\frac{1}{q^6}\ g^{\alpha\beta}:\left\{\bar{q}^i \left(S_{\mu\alpha}  -
A_{\mu\alpha}\right) Q\,
\psi^j
\,\, \bar{q}^j\left({S}_{\nu\beta} -
{A}_{\nu\beta}
\right)
T_3\,
\gamma_{5}\psi^{i}\right\}(0):
\nn\\
\cU_{\mu\nu}^{(2)}(q){\Big\vert}_{q\ra\infty}
& = &  -i\frac{i^2}{2!} g_s^2
\frac{1}{q^6}\ g^{\alpha\beta}:\left\{\bar{\psi}^i \left(S_{\mu\alpha}
+ A_{\mu\alpha}\right) Q\,q^j
\,\, \bar{\psi}^j\left({S}_{\nu\beta} +
{A}_{\nu\beta}
\right)
T_3\,\gamma_{5}q^{i}\right\}(0):\nn \\
\cU_{\mu\nu}^{(3)}(q){\Big\vert}_{q\ra\infty}
& = &  +i\frac{i^2}{2!} g_s^2
\frac{1}{q^6}\ g^{\alpha\beta}:\left\{\bar{q}^i \left(S_{\mu\alpha}  -
A_{\mu\alpha}\right) Q\,
\psi^j
\,\, \bar{\psi}^j\left({S}_{\nu\beta} +
{A}_{\nu\beta}
\right)
T_3\,
\gamma_{5}q^{i}\right\}(0):\nn\\
\cU_{\mu\nu}^{(4)}(q){\Big\vert}_{q\ra\infty}
& = &  +i\frac{i^2}{2!} g_s^2
\frac{1}{q^6}\ g^{\alpha\beta}:\left\{\bar{\psi}^i \left(S_{\mu\alpha}
+  A_{\mu\alpha}\right) Q\,
q^j
\,\, \bar{q}^j\left({S}_{\nu\beta} -
{A}_{\nu\beta}
\right)
T_3\,
\gamma_{5}\psi^{i}\right\}(0):\nn\,
\eea}

\noi
where $S_{\mu\alpha}$ and  $A_{\mu\alpha}$
 denote the tensors:
$$
S_{\mu\alpha}  =  {q}_\mu \gamma_\alpha +{q}_\alpha\gamma_\mu-
g_{\mu\alpha}\not\! q\,,\qquad\annd\qquad
A_{\mu\alpha}=iq^\eta\epsilon_{\mu\eta\alpha\sigma}\gamma^\sigma
\gamma_5\,.
$$

\noi
Only the antisymmetric tensors contribute in the total sum, with the
result
$$
\cU_{\mu\nu}(q){\big\vert}_{q \ra\infty}
= -i\frac{i^2}{2!} g_s^2\frac{1}{q^6}g^{\alpha\beta}
4(i)^2 q^\eta q^\lambda\,
\epsilon_{\mu\eta\alpha\sigma}\,\epsilon_{\nu\lambda\beta\rho}
(\bar{q}^i\gamma^\sigma \gamma_5 Q\psi^j)(\bar{q}^j \gamma^\rho T_3
\psi^i)(0)\,.
$$
Fierzing the resulting four--quark operator, using the relevant terms of
the identity
$$
 (\bar{q}^i   \gamma^\alpha \gamma_5  q^j)( \bar{\psi}^j\gamma^\beta
\psi^i) = -\frac{1}{8} \epsilon^{\alpha\beta \rho\sigma}
\big[(\bar{q}\psi)(\bar{\psi}\sigma_{\rho\sigma} q) +
(\bar{q}\sigma_{\rho\sigma}\psi)(\bar{\psi} q) \big] \, + \, \cdots
\,,
$$
and contracting two of the epsilon tensors (using the correct sign!)
one finally gets the result
$$
\cU_{\mu\nu}(q)=i\left[q_{\delta}
\epsilon_{\mu\nu\alpha\beta}q^{\alpha}-
q_{\beta}\epsilon_{\mu\nu\alpha\delta}q^{\alpha}\right]
\left(-2\pi^2\frac{\als}{\pi}\right)\frac{\cO^{\beta\delta}(0)}{\left(Q^2
\right)^3}+\cdots\,;
$$
with $\cO^{\beta\delta}(0)$ the tensor
$$
\cO^{\beta\delta}(0)=\left[\frac{2}{3}\left( \bar{u}\sigma^{\beta\delta}
u\right)\left(\bar{u}u \right)+\frac{1}{3}\left(
\bar{d}\sigma^{\beta\delta} d\right)\left(\bar{d}d
\right)+\frac{1}{3}\left( \bar{s}\sigma^{\beta\delta}
s\right)\left(\bar{s}s \right)\right](0)\,.
$$


\vspace*{1cm}

\section{\large The OPE beyond the Chiral Limit}
\setcounter{equation}{0}
\def\theequation{\Alph{section}.\arabic{equation}}

In the presence of quark masses, there are two types of terms that one
has to consider in the evaluation of the function in Eq.~\rf{Ufunction}.
(Recall that the relevant terms in the OPE are those with the explicit
tensor structure shown in Eq.~\rf{opeterm}.)
There are terms of $\cO(m)$ from the expansion of the quark propagator:
$$
\frac{1}{\qs-m}=\frac{1}{\qs}+\frac{1}{\qs}m\frac{1}{\qs}+\cO(m^2)\,,
$$
and terms from the Taylor expansion of the quark field
$$
q(x)=q(0)+x^{\sigma}D_{\sigma}q(0)+\cdots\,.
$$

The first type expansion leads to a potential contribution:
\be\lbl{propag}
\cU_{\mu\nu}(q)\big\vert_{m}= i\
\epsilon_{\mu\nu\alpha\beta}\
\frac{1}{Q^2}\ \bar{q}(0)\sigma^{\alpha\beta} \cM QQ_{L}^{(3)}q(0)\,,
\ee
where $\cM$ denotes the quark mass matrix
\be
\cM=\mbox{\rm diag} (m_u, m_d, m_s)\,.
\ee
This contribution, however, does not have the two powers of
$q$--momentum required by the tensor structure in Eq.~\rf{opeterm}.

The Taylor expansion from the quark field leads to the operator
\be\lbl{taylorq}
\cU_{\mu\nu}(q)\big\vert_{q}=D_{\rho}\bar{q}(0)\gamma_{\mu}
\frac{\gamma^{\rho}q^2-2\qsls q^{\rho}}{Q^4}
\gamma_{\nu}\gamma_{5}QQ_{L}^{(3)}q(0)-\bar{q}(0)\gamma_{\nu}\gamma_{5}
\frac{\gamma^{\rho}q^2-2\qsls q^{\rho}}{Q^4}
\gamma_{\mu} Q_{L}^{(3)}Q\ D_{\rho}q(0)\,.
\ee
Using the Dirac equation
\be\lbl{dirac}
(iD^{\rho}\gamma_{\rho}-\cM)q=0\,,
\ee
the terms proportional to
$\gamma^{\rho}$ in Eq.~\rf{taylorq} lead to the operator
\be\lbl{gammarho}
\cU_{\mu\nu}(q)\big\vert_{\gamma^{\rho}}=-\frac{2}{Q^2}\left[
D_{\mu}\bar{q}(0)\gamma_{\nu}\gamma_{5} QQ_{L}^{(3)}q(0)
-\bar{q}(0)\gamma_{\nu}\gamma_{5}
QQ_{L}^{(3)}D_{\mu}q(0)\right]\,,
\ee
where we have used the fact that terms
$\sim \sigma^{\alpha\beta}$ cancel with those from the quark--propagator
contribution in Eq.~\rf{propag}. The other terms in Eq.~\rf{taylorq} are
proportional to
$\qsls$ , and lead to the following combination of
operators
{\setl
\bea
\cU_{\mu\nu}(q)\big\vert_{\not q} & =  &
-\frac{2}{Q^4}\left\{\ iq_{\beta}
\epsilon_{\mu\nu\alpha\delta}q^{\alpha}
\left(D^{\beta}
\bar{q}(0)\gamma^{\delta}QQ_{L}^{(3)}q(0)+
\bar{q}(0)\gamma^{\delta}QQ_{L}^{(3)}D^{\beta}q(0)\right)\right. \nn \\
 & &\left. +
q^{\beta}(q_{\mu}\delta_{\delta}^{\beta}-g_{\mu\nu}q^{\delta}+
q_{\nu}\delta_{\mu}^{\delta})
\left(D_{\beta}\bar{q}(0)\gamma_{\delta}\gamma_{5}QQ_{L}^{(3)}q(0)-
\bar{q}(0)\gamma_{\delta}\gamma_{5}QQ_{L}^{(3)}D_{\beta}q(0)\right)\right\}\,.
\eea}

When inserting the operators $\cU_{\mu\nu}(q)\big\vert_{\gamma^{\rho}}$
and
$\cU_{\mu\nu}(q)\big\vert_{\not q}$ in the $T$--product in
the r.h.s. of Eq.~\rf{VAV} there appear the following two--point functions
\be\lbl{xitpf}
\Xi^{\beta\delta}_{~~\rho}(k)=\int d^4 y e^{-ik\cdot y}
\langle\Omega\vert T\left\{
\left(D^{\beta}\bar{q}\gamma^{\delta}QQ_{L}^{(3)}q+
\bar{q}\gamma^{\delta}QQ_{L}^{(3)}D^{\beta}q
\right)(0) V_{\rho}^{\elm}(y) \right\}\vert\Omega\rangle \,,
\ee
and
\be\lbl{deltatpf}
\Delta_{\beta\delta\rho}(k)=\int d^4 y e^{-ik\cdot y}
\langle\Omega\vert T\left\{
\left(D_{\beta}\ \bar{q}\gamma_{\delta}\gamma_{5}QQ_{L}^{(3)}q-
\bar{q}\gamma_{\delta}\gamma_{5}QQ_{L}^{(3)}D_{\beta}q\right)(0)
V_{\rho}^{\elm}(y) \right\}\vert\Omega\rangle \,.
\ee
Lorentz covariance and gauge invariance of the electromagnetic
current restrict these two--point functions to the following
decomposition~\footnote{There is in fact a second possible term in
$\Xi_{\beta\delta\rho}(k)$ of the type
$-(g_{\beta\rho}k^2-k_{\beta}k_{\rho})\ k_{\delta}
\ \tilde{\Pi}_{V'V}(k^2)$, which however vanishes   as can be seen
by contracting the $\beta$ and $\rho$ indices in the r.h.s of
Eq.~\rf{xitpf}, using translation invariance and conservation of
$V_{\rho}^{\mbox{\rm\tiny em}}$.}
\be\lbl{vv}
\Xi_{\beta\delta\rho}(k)=-(g_{\delta\rho}k^2-k_{\delta}k_{\rho})\
k_{\beta}
\Pi_{V'V}(k^2)\,,
\ee
and
\be\lbl{av}
\Delta_{\beta\delta\rho}(k)=i\epsilon_{\beta\rho\delta\alpha}k^{\alpha}
\Pi_{A'V}(k^2)\,.
\ee
We can now obtain the asymptotic behaviour of the Green's function
$W_{\mu\nu\rho}(q,k)$ defined  in Eq.~\rf{VAV} in terms of the invariant
functions $\Pi_{V'V}(k^2)$ and $\Pi_{A'V}(k^2)$, with the result
{\setl
\bea
\lim_{q\ra\infty}W_{\mu\nu\rho}(q,k) & = & -\frac{2i}{Q^2}
\epsilon_{\mu\rho\nu\alpha}k^{\alpha}\Pi_{A'V}(k^2) \nn \\
 & & -\frac{2i}{Q^4}\left\{
\left(q_{\mu}\epsilon_{\beta\rho\nu\alpha}+
q_{\nu}\epsilon_{\beta\rho\mu\alpha}
\right)q^{\beta}k^{\alpha}\Pi_{A'V}(k^2)
\right.
\nn
\\ \lbl{opemass}
 & & \left.+\epsilon_{\mu\alpha\nu\beta}\left(\delta_{\rho}^{\beta}k^2
-k^{\delta}k_{\beta}\right)q^{\alpha}\ (q\cdot k)\Pi_{V'V}(k^2)\right\}\,.
\eea}

The next step consists in extracting from the expression in
Eq.~\rf{opemass} the transverse component:
$\lim_{q\ra\infty}W_{\mu\nu\rho}^{\mbox{\rm\tiny trans}}(q,k)$, defined in
Eq.~\rf{Wk}. Taking the derivative with respect to the $k$--momentum,
as defined in Eq.~\rf{Wk}, we finally get
$$
\lim_{q\ra\infty}W_{\mu\nu\rho\sigma}(q)=-\frac{2i}{Q^4}
\left[q_{\rho}\epsilon_{\mu\nu\alpha\sigma}
q^{\alpha}-q_{\sigma}\epsilon_{\mu\nu\alpha\rho}q^{\alpha}
\right]\Pi_{A'V}(0)\,.
$$
It follows then that the asymptotic behaviour of the function
$W(Q^2)$, in the presence of quark masses, is given by
$$
\lim_{Q^{2}\ra\infty}W(Q^2)=-\frac{2}{Q^4}\Pi_{A'V}(0)\,.
$$

What do we know about $\Pi_{A'V}(0)$? It turns out that the invariant
functions $\Pi_{A'V}(k^2)$, $\Pi_{V'V}(k^2)$ defined in Eqs.~\rf{vv},
\rf{av} and the invariant function $\Pi_{VT}(k^2)$ defined in
Eq.~\rf{VTfunc} are not independent. This can be seen starting with the
identity
$$
i\bar{q}\sigma_{\alpha\beta}\cM QQ_{L}^{(3)}q=
\frac{1}{2}\left(D^{\rho}\bar{q}\gamma_{\rho}\sigma_{\alpha\beta}
QQ_{L}^{(3)}q-
\bar{q}\sigma_{\alpha\beta}\gamma_{\rho}QQ_{L}^{(3)}D^{\rho}q \right)\,,
$$
which follows from the Dirac equation \rf{dirac}, and using the Dirac
algebra identity in Eq.~\rf{diracid}
which brings in the relevant operators. The resulting relation, to lowest
order in the quark masses, is
$$
2\tr\left(\cM QQ_{L}^{(3)}Q\right)\Pi_{VT}(k^2)=
\frac{1}{2}\ k^{2}\Pi_{V'V}(k^2)+\Pi_{A'V}(k^2)\,,
$$
which at $k^2=0$ reduces to
$$
\Pi_{A'V}(0)=\frac{2}{9}(4m_{u}-m_{d}-m_{s})\Pi_{VT}(0)\,.
$$


\vspace*{1cm}

\section{\large On the presence of $\log M_Z$ terms in
$a_{\mu}^{EW(2)}(e,\mu,u,d,s,c)$}
\setcounter{equation}{0}
\def\theequation{\Alph{section}.\arabic{equation}}

We have noticed in section {\bf 7.2} that for the two first
generations, the $\log M_Z$ terms do not cancel in
$a_{\mu}^{EW(2)}(e,\mu,u,d,s,c)$, see Eq. \rf{12g}. This result
might appear as puzzling, since in the case of free quarks, which
encompasses the constituent quark model, the cancellation of the
$\log M_Z$ terms in $a_{\mu}^{EW(2)}(e,\mu,u,d,s,c)$ occurs, as
observed in ref. \cite{CKM95}, and as shown in Eq. \rf{12gCKM}.
However, this latter situation is conceptually very different from
the one described by {\it confined} light (or even massless)
quarks. This difference, that needs perhaps to be explained more
in detail, is at the origin of the non-cancellation of the $\log
M_Z$ contributions in Eq. \rf{12g}.

\indent

In order to proceed, let us start from Eq. \rf{WchiQM}, that we write
in a slightly different, but more convenient, form,

{\setl \bea w_{\mbox{$\chi$\rm\footnotesize QM}}(z) & = &
\frac{N_c}{12\pi^2}\frac{8}{3} g_{A}\frac{M_{\rho}^2}{M_{Q}^2}
\times \int_{0}^{1}dy \int_{0}^{1-y}dx
\frac{xy-y(1-y)}{1+\frac{Q^2}{M_{Q}^2} y(1-y)} \nn \\
& = & \int_{-\infty}^{\infty} dt\,\frac{1}{t + Q^2-i \varepsilon
}\,\frac{1}{\pi} \mathrm{Im} \, w_{\mbox{$\chi$\rm\footnotesize
QM}}(t) \,. \eea} \noindent The second,i.e. a dispersive form, is
obtained upon performing the integration over the Feynman
parameter $x$ in the expression on the first line, followed by an
obvious change of variable in the remaining integration over $y$,
with the outcome \bea \frac{1}{\pi}\mathrm{Im} \,
w_{\mbox{$\chi$\rm\footnotesize QM}}(t) \,=\,
-\,\frac{1}{2}\,\frac{M_Q^2}{t^2}\,\frac{1}{\sqrt{1-\frac{4M_Q^2}{t}}}
\,\theta(t-4M_Q^2)\,\left(\frac{N_c}{12\pi^2}\frac{8}{3}
g_{A}{M_{\rho}^2}\right)\,. \eea
For the case of a massless free fermion, one obtains \bea
\lim_{M_Q\to 0}w_{\mbox{$\chi$\rm\footnotesize QM}}(z) \, =
\,\frac{-N_c}{12\pi^2}\frac{2}{3} g_{A}\frac{M_{\rho}^2}{Q^2-i
\varepsilon} \,, \eea since, as is well known, \bea \lim_{M_Q\to
0}-\,\frac{1}{2}\,\frac{M_Q^2}{t^2}\,
\frac{1}{\sqrt{1-\frac{4M_Q^2}{t}}}\,\theta(t-4M_Q^2) \, =\,
-\,\frac{1}{4}\,\delta(t)\,. \eea
Thus, in the massless limit, the function
$w_{\mbox{$\chi$\rm\footnotesize QM}}(z)$ reduces to a simple
pole. This singularity at $Q^2=0$ also provides the $1/Q^2$
behaviour at large $Q^2$, which in turn will lead to the
cancellation of the $\log M_Z$ terms. Notice also that this
asymptotic $1/Q^2$ behaviour, and the ensuing cancellation of
$\log M_Z$ terms, is preserved for a massive fermion, as shown by
Eq. \rf{WchiQM} or by the dispersive representation above.

\indent

The real question we now have to address is how much of all this
can be taken over to QCD. Let us first discuss the chiral limit.
One immediate and important observation one has to make in the
case of QCD is that the pole we have just discussed in the case
of a massless free fermion no longer occurs. Indeed, one may
evaluate the relevant three-point function
$W_{\mu\nu\rho}(q_1,q_2)$ within the effective low-energy chiral
lagrangian. The lowest order contribution starts at order ${\cal
O}(p^4)$ and is given by the anomaly pole, as discussed in ref.
\cite{PPdeR95}. At NLO, ${\cal O}(p^6)$ in this case, we have a
counterterm contribution~\footnote{We stay of within the
large-$N_C$ framework, hence meson loops can be ignored.}, which
leads to a non vanishing, but {\it constant}, value of
$w(Q^2)$~\footnote{In terms of the ${\cal O}(p^6)$ chiral
Lagrangian constructed for the odd intrinsic parity sector in
\cite{Bijnensetal02}, the relevant low-energy constant is
$C_{22}^W$.}. Thus, confinement drastically modifies the long
distance behaviour of $W(Q^2)$ as compared to the free fermion
case, not really a surprise. It now remains to discuss the short
distance aspects. We may write the function $w(Q^2)$ in the
dispersive form given before for the free fermion case, \bea
w(Q^2) \, = \, \int_{-\infty}^{\infty} dt\,\frac{1}{t + Q^2-i
\varepsilon }\,\frac{1}{\pi} \mathrm{Im} \, w(t) \,, \eea but
where now the perturbative QCD part describes the absorbtive part
above a certain threshold $s_0$, which has to be sufficiently
large so that perturbative QCD can be applied. Within the
large-$N_C$ framework, we may thus write \bea
\frac{1}{\pi}\mathrm{Im} \, w(t) \,=\, \frac{1}{\pi}\mathrm{Im}
\, w^{Res}(t) \,+\, \frac{1}{\pi}\mathrm{Im} \, w^{pQCD}(t) \,.
\eea The non perturbative piece from the narrow width resonant
states writes \bea \frac{1}{\pi}\mathrm{Im} \, w^{Res}(t) \, =
\,\sum_{R}M_{\rho}^2{\alpha_{R}}\delta({t-M_{R}^2})\,, \eea and
$M_{R}$ the mass of the $R$ narrow state, $M_{R}^2 < s_0$. The
perturbative contribution for a massive light quark with a {\rm
current algebra} mass $m_q$, to lowest order in perturbative QCD
(free quarks, $g_A=1$), is \bea \frac{1}{\pi}\mathrm{Im} \,
w^{pQCD}(t) \,= \, -
\frac{1}{2}\,\frac{m_q^2}{t^2}\,\frac{1}{\sqrt{1-\frac{4m_q^2}{t}}}
\,\theta(t-s_0)\,\left(\frac{N_c}{12\pi^2}\frac{8}{3} g_A
{M_{\rho}^2}\right)\,. \eea
This gives then
\bea
w(z) \, = \, w^{Res}(z)\,+\,w^{pQCD}(z)
\,,
\eea
with
\bea
w^{Res}(z) &=& \sum_{R}\frac{\alpha_{R}}{z+z_{R}}\,,\quad
z_{R}=\frac{M_{R}^2}{M_{\rho}^2}\,,
\nn\\
w^{pQCD}(z) \!\!\!\!&=&
\!\!\!\!\!-\,\frac{1}{4}\,\frac{M_\rho^2}{Q^2}
\left[1\,-\,\sqrt{1-\frac{4m_q^2}{s_0}}\,
-\,\frac{2m_q^2}{Q^2}\,\log\left( 1+\frac{Q^2}{s_0}\right) + {\cal
O}(m_q^4)\right]\! \left(\frac{N_c}{12\pi^2}\frac{8}{3}g_A
\right)\left(1+{\cal O}(\alpha_s)\right) \,. \nn \eea We see here
the crucial difference with the constituent quark model, where
$s_0$ would coincide with $4M_Q^2$, thus providing the same
leading $1/Q^2$ short distance behaviour as for a massless free
quark. In QCD, the onset of the continuum occurs at a scale $s_0$
that is much larger than the scale set by the light quark masses,
and for $s_0\gg 4m_q^2$, one finds \bea \lim_{z\to\infty} z
w^{pQCD}(z) \,=\, {\cal O}(m_q^2) \,. \nn \eea Thus, in the
chiral limit, or if one considers only first order explicit
chiral symmetry breaking effects, the short distance behaviour of
$w(z)$ in QCD reads as given in Eq. \rf{massOPE}, \bea
\lim_{z\to\infty}w(z)=-\frac{4}{9} \frac{(4m_{u}-m_{d}-m_{s})
\stern_{0}}{Q^4}+
\frac{16}{9}\pi^2\frac{\als}{\pi}\frac{\stern_{0}^2}{Q^6}+\cdots\,,
\nn \,. \eea The fact that the leading short distance fall off of
$w(z)$ in QCD is $1/z^3$ in the chiral limit, or $1/z^2$ for
massive light quark if only first order order explicit chiral
symmetry breaking effects are retained, explains why the
cancellation of $\log M_Z$ terms no longer occurs in QCD.

\indent

We conclude therefore that the cancellation
of the $\log M_Z$ terms does not occur in QCD in the chiral limit,
and by stressing once again very strongly that constituent
quark models do not, in general, provide an adequate description
of fundamental properties of QCD.

\vspace{0.5cm}



\begin{thebibliography}{99}

\bibitem{BNL}
         H.N.~Brown {\it et al.}, Phys. Rev. Lett. {\bf 86}
         (2001) 2227.

\bibitem{CM01}
         A.~Czarnecki and W.J.~Marciano, Phys. Rev. {\bf D64} (2001)
         013014.

\bibitem{DH98}
         M.~Davier and A.~H\"{o}cker, Phys. Lett. {\bf B435} (1998) 427.

\bibitem{J00}
         F.~Jegerlehner, hep-ph/0104304.

\bibitem{N01}
         S.~Narison, Phys. Lett. {\bf B513} (2001) 53; Erratum {\it
         ibid.} {\bf B526} (2002) 414.

\bibitem{DTY01}
         J.F.~de Troc\'oniz and F.J.~Yndur\'ain, Phys. Rev. {\bf D65}
         (2002) 093001.

\bibitem{CLS01}
         G.~Cveti\v{c}, T.~Lee and I.~Schmidt, Phys. Lett. {\bf B520} (2001)
         222.

\bibitem{KNb01}
         M.~Knecht and A.~Nyffeler, Phys. Rev. {\bf D65} (2002) 073034.

\bibitem{KNO85}
         T.\ Kinoshita, B.\ Ni\v zi\'c and Y.\ Okamoto,  Phys. Rev. Lett.
         {\bf 52} (1984) 717;  Phys.\ Rev.\ {\bf D31}, (1985) 2108.

\bibitem{HKS9596}
         M.~Hayakawa, T.~Kinoshita and A.~I.~Sanda, Phys.\ Rev.\ Lett.\
         {\bf 75} (1995) 790;  M.\ Hayakawa and T.\ Kinoshita,
         Phys.\ Rev.\ {\bf D57} (1998) 465.

\bibitem{BPP}
         J.\ Bijnens, E.\ Pallante and J.\ Prades, Phys.\ Rev.\
         Lett.\ {\bf 75} (1995) 1447; Erratum {\it ibid.}, {\bf 75} (1995)
         3781; Nucl.\ Phys.\ {\bf B474} (1996) 379 .

\bibitem{BP01}
         J. Bijnens and F. Persson, hep-ph/0106130.

\bibitem{bartos01}
         E. Barto{\v s}, A.Z.~Dubnickova, S.~Dubnicka, E.A.~Kuraev and
         E.~Zemlyanaya, hep-ph/0106084.


\bibitem{KNPdeR01}
         M.~Knecht, A.~Nyffeler, M.~Perrottet and E.~de Rafael,
         Phys. Rev. Lett. {\bf 88} (2002) 071802 .

\bibitem{Ba01}
         W.~Bardeen, {\it private communication.}

\bibitem{BCM01}
         I.~Blockland, A.~Czarnecki and K.~Melnikov, Phys. Rev. Lett.
         {\bf 88} (2002) 071803.

\bibitem{RW02}
         M.J.~Ramsey-Musolf and M.~Wise, hep-ph/0201297v2.

\bibitem{HK01}
         M.~Hayakawa and T.~Kinoshita, hep-ph/0112102.

\bibitem{BPPe01}
         J.\ Bijnens, E.\ Pallante and J.\ Prades, Nucl. Phys. {\bf B626}
         (2002) 410.

\bibitem{bartosR01}
         E. Barto{\v s}, A.Z.~Dubnickova, S.~Dubnicka, E.A.~Kuraev and
         E.~Zemlyanaya,, hep-ph/0106084v2.

\bibitem{KKSS92}
         T.V.~Kukhto, E.A.~Kuraev, A.~Schiller and
         Z.K.~Silagadze, Nucl. Phys. {\bf B371} (1992) 567.

\bibitem{PPdeR95}
         S.~Peris, M.~Perrottet and E.~de Rafael, Phys. Lett.
         {\bf B355} (1995) 523.

\bibitem{CKM95}
         A.~Czarnecki, B.~Krause and W.~Marciano, Phys. Rev. {\bf D52}
         (1995) R2619.

\bibitem{BBdeR93}
        J. Bijnens, C. Bruno and E. de Rafael, Nucl. Phys. {\bf
        B390} (1993) 389. For a critical review and update of
        this type of models see S. Peris, M. Perrottet and E. de
        Rafael, JHEP 9805 (1998) 011.

\bibitem{deR01}
         E.~de Rafael, {\it Large--$N_c$ QCD and Low Energy
         Interactions}, hep-ph/0110195.

\bibitem{KPdeR98}
         M.~Knecht, S.~Peris and E.~de Rafael, Phys. Lett. {\bf B443}
         (1998) 255.

\bibitem{KPPdeR99}
         M.~Knecht, S.~Peris, M.~Perrottet and E.~de Rafael, Phys.
         Rev. Lett. {\bf 83} (1999) 5230.

\bibitem{PdeR00}
         S.~Peris and E.~de Rafael, Phys. Lett. {\bf B490} (2000) 213;
         Erratum, hep-ph/0006146v3.

\bibitem{GP00}
         M. Golterman and S. Peris, Phys. Rev. \textbf{D61} (2000)
         034018.

\bibitem{KPdeR01}
         M.~Knecht, S.~Peris and E.~de Rafael, Phys. Lett. {\bf B508}
         (2001) 117.

\bibitem{SV}
         S. Narison, G.~Shore and G. Veneziano, Nucl. Phys. {\bf
         B391} (1993) 69.

\bibitem{KN01}
         M.~Knecht and A.~Nyffeler, Eur. Phys. J. {\bf C21} (2001) 659.

\bibitem{WW}
         E.T.~Whittaker and G.N.~Watson, {\it A course of MODERN
         ANALYSIS} Cambridge University Press, 4th edition 1965.

\bibitem{KdeR98}
         M.~Knecht and E.~de Rafael, Phys. Lett. {\bf B424} (1998) 335.

\bibitem{EGLPdeR89}
         G.~Ecker, J.~Gasser, H.~Leutwyler, A.~Pich and E.~de Rafael,
         Phys. Lett. {\bf B223} (1989) 311.

\bibitem{BGL72}
         W.A.~Bardeen, R.~Gastmans and B.E.~Lautrup, Nucl. Phys. {\bf
         B46} (1972) 315.

\bibitem{ACM72}
         G.~Altarelli, N.~Cabbibo and L.~Maiani, Phys. Lett. {\bf 40B}
         (1972) 415.

\bibitem{JW72}
         R.~Jackiw and S.~Weinberg, Phys. Rev. {\bf D5} (1972) 2473.

\bibitem{BY72}
         I.~Bars and M.~Yoshimura, Phys. Rev. {\bf D6} (1972) 374.

\bibitem{FLS72}
         M.~Fujikawa, B.W.~Lee and A.I.~Sanda, Phys. Rev. {\bf D6}
         (1972) 2923.

\bibitem{Smi94}
         V.A.~Smirnov, Mod. Phys. Lett.{ \bf A10} (1995) 1485.

\bibitem{CKM96}
         A.~Czarnecki, B.~Krause and W.J.~Marciano, Phys. Rev. Lett.
        {\bf 76} (1996) 3267.

\bibitem{DG98}
         G.~Degrassi and G.F.~Giudice, Phys. Rev. {\bf D58} (1998)
         053007.

\bibitem{NSVZ84}
         V.A.~Novikov, M.A.~Shifman, A.I.~Vainshtein and V.I.~Zakharov,
         Fortschr. Phys. {\bf 32} (1984) 585.

\bibitem{Bijnensetal02}
J. Bijnens, L. Girlanda and P. Talavera, Eur. Phys. J. C {\bf 23}
(2002) 539.


\end{thebibliography}
\end{document}